\documentclass[twocolumn]{aastex631}

\newcommand{\ha}{H$\alpha$}
\newcommand{\hb}{H$\beta$}

\def\tarq{GNz7q}
\def\targ{ND1}

\newcommand{\mgii}{Mg\textsc{ii}$\lambda\lambda$2796,2803}

\newcommand{\oi}{\textsc{Oi}$\lambda$1302}

\newcommand{\siii}{Si\textsc{ii}$\lambda$1260}

\newcommand{\ciiemi}{{\textsc{[C\,ii]}}}
\newcommand{\oiiiemi}{{\textsc{[O\,iii]}}}

\newcommand{\cigale}{\texttt{cigale}}
\usepackage{hyperref}

\shorttitle{Dense, metal-enriched CGM at $z= 7$}
\shortauthors{Fei et al.}
\graphicspath{{./}{figures/}}

\usepackage{amsmath}
\usepackage{multirow}

\begin{document}

\title{
A dense, metal-rich absorber driven by a heavily dust-obscured galaxy:

The direct evidence of CGM enrichment at $z\sim7$
}

\author[0000-0001-7232-5355]{Qinyue Fei}
\affiliation{David A. Dunlap Department of Astronomy and Astrophysics, University of Toronto, 50 St. George Street, Toronto, Ontario, M5S 3H4, Canada}
\email{qy.fei@utoronto.ca}

\author[0000-0001-7201-5066]{Seiji Fujimoto}
\affiliation{David A. Dunlap Department of Astronomy and Astrophysics, University of Toronto, 50 St. George Street, Toronto, Ontario, M5S 3H4, Canada}
\affiliation{Dunlap Institute for Astronomy and Astrophysics, 50 St. George Street, Toronto, Ontario, M5S 3H4, Canada}

\author[0000-0003-2680-005X]{Gabriel Brammer}
\affiliation{Cosmic Dawn Center (DAWN), University of Copenhagen, Denmark}
\affiliation{Niels Bohr Institute, University of Copenhagen, Jagtvej 128, DK-2200 N, Copenhagen, Denmark}

\author[0000-0001-8415-7547]{Lise Christensen}
\affiliation{Cosmic Dawn Center (DAWN), University of Copenhagen, Denmark}
\affiliation{Niels Bohr Institute, University of Copenhagen, Jagtvej 128, DK-2200 N, Copenhagen, Denmark}

\author[0000-0001-9191-9837]{Marianne Vestergaard}
\affiliation{DARK, Niels Bohr Institute, University of Copenhagen, Jagtvej 128, DK-2200 N, Copenhagen, Denmark}
\affiliation{Steward Observatory, University of Arizona, 933 N Cherry Avenue, Tucson, AZ 85721, USA}

\author[0000-0003-3997-5705]{Rohan P.~Naidu}\altaffiliation{Hubble Fellow}
\affiliation{
MIT Kavli Institute for Astrophysics and Space Research, 70 Vassar Street, Cambridge, MA 02139, USA}

\author[0000-0003-3769-9559]{Robert A. Simcoe}
\affiliation{MIT Kavli Institute for Astrophysics and Space Research, 70 Vassar Street, Cambridge, MA 02139, USA}
\affiliation{Department of Physics, Massachusetts Institute of Technology, Cambridge, MA 02139, USA}

\author[0000-0002-8659-3729]{Norman Murray}
\affiliation{Canadian Institute for Theoretical Astrophysics, University of Toronto, 60 St. George Street, Toronto, ON M5S 3H8, Canada}

\author[0000-0001-6947-5846]{Luis C. Ho}
\affiliation{Kavli Institute for Astronomy and Astrophysics, Peking University, Beijing 100871, China}
\affiliation{Department of Astronomy, School of Physics, Peking University, Beijing 100871, China}

\author[0000-0001-8496-4162]{Ruancun Li}
\affiliation{Max-Planck-Institut f{\"u}r extraterrestrische Physik, Gie{\ss}enbachstra{\ss}e 1, 85748 Garching bei M{\"u}nchen, Germany}

\author[0000-0003-0212-2979]{Volker Bromm}
\affiliation{Department of Astronomy, University of Texas, Austin, TX 78712, USA}
\affiliation{Weinberg Institute for Theoretical Physics, University of Texas, Austin, TX 78712, USA}
\affiliation{Cosmic Frontier Center, The University of Texas at Austin, Austin, TX 78712, USA}

\author[0000-0002-7093-1877]{Javier \'{A}lvarez-M\'{a}rquez}
\affiliation{Centro de Astrobiolog\'{\i}a (CAB), CSIC-INTA, Ctra. de Ajalvir km 4, Torrej\'{o}n de Ardoz, E-28850, Madrid, Spain}

\author[0000-0003-3596-8794]{Hollis B. Akins}
\altaffiliation{NSF Graduate Research Fellow}
\affiliation{Department of Astronomy, The University of Texas at Austin, Austin, TX 78712, USA}

\author[0000-0003-3983-5438]{Yoshihisa Asada}
\affiliation{David A. Dunlap Department of Astronomy and Astrophysics, University of Toronto, 50 St. George Street, Toronto, Ontario, M5S 3H4, Canada}
\affiliation{Dunlap Institute for Astronomy and Astrophysics, 50 St. George Street, Toronto, Ontario, M5S 3H4, Canada}

\author[0009-0009-6673-0851]{Simona Di Stefano}
\affiliation{Dipartimento di Fisica, Sezione di Astronomia, Università di Trieste, Via Tiepolo 11, I-34143 Trieste, Italy}
\affiliation{INAF, Osservatorio Astronomico di Trieste, Via Tiepolo 11, I-34143 Trieste, Italy}

\author[0000-0000-0000-0001]{Kohei Inayoshi}
\affiliation{Kavli Institute for Astronomy and Astrophysics, Peking University, Beijing 100871, China}

\author[0000-0002-5588-9156]{Vasily Kokorev}
\affiliation{Department of Astronomy, University of Texas, Austin, TX 78712, USA}
\affiliation{Cosmic Frontier Center, The University of Texas at Austin, Austin, TX 78712, USA}

\author[0000-0003-2871-127X]{Jorryt Matthee}
\affiliation{Institute of Science and Technology Austria (ISTA), Am Campus 1, 3400 Klosterneuburg, Austria}

\author[0000-0001-5492-4522]{Romain A. Meyer}
\affiliation{Department of Astronomy, University of Geneva, Chemin Pegasi 51, 1290 Versoix, Switzerland}

\author[0000-0002-5283-933X]{Ava Polzin}
\affiliation{Department of Astronomy \& Astrophysics, The University of Chicago, Chicago, IL 60637 USA}
\affiliation{Kavli Institute for Cosmological Physics, The University of Chicago, Chicago, IL 60637 USA}

\author[0000-0001-6477-4011]{Francesco Valentino }
\affiliation{Cosmic Dawn Center, DTU Space, Technical University of Denmark, Elektrovej 327, DK-2800 Kgs. Lyngby, Denmark}

\author[0000-0003-4793-7880]{Fabian Walter}
\affiliation{Max Planck Institut f\"ur Astronomie, K\"onigstuhl 17, D-69117 Heidelberg, Germany}
\affiliation{California Institute of Technology, Pasadena, CA 91125, USA}
\affiliation{National Radio Astronomy Observatory, Pete V. Domenici Array Science Center, P.O. Box O, Socorro, NM 87801, USA}

\author[0000-0002-9838-8191]{Marcel Neeleman}
\affiliation{National Radio Astronomy Observatory, 520 Edgemont Road, Charlottesville, VA 22903, USA}

\author[0000-0003-0111-8249]{Yunjing Wu}
\affiliation{Kavli Institute for the Physics and Mathematics of the Universe (WPI), The University of Tokyo Institutes for Advanced Study, The University of Tokyo, Kashiwa, Chiba 277-8583, Japan}
\affiliation{Center for Data-Driven Discovery, Kavli IPMU (WPI), UTIAS, The University of Tokyo, Kashiwa, Chiba 277-8583, Japan}

\author[0000-0003-1207-5344]{Mengyuan Xiao}
\affiliation{Department of Astronomy, University of Geneva, Chemin Pegasi 51, 1290 Versoix, Switzerland}

\begin{abstract}
We report the serendipitous discovery of a dense, metal-enriched circumgalactic medium (CGM) absorber at $z=7.03$ and its host galaxy along the sightline toward a red quasar. 
The absorber shows the largest rest-frame equivalent widths ($\approx$2--4 times those of the strongest $z>6$ absorbers in previous surveys) and column densities ($\gtrsim$1--1.5 dex higher than the median of known $z>5$ metal absorbers) observed to date at $z>6$, tracing a metal-enriched gaseous halo. With deep JWST/NIRSpec and NOEMA observations, we identify the host galaxy, ND1, as a star-forming galaxy (SFR $\approx$ 26-170 $M_\odot$ yr$^{-1}$) at an impact parameter of $\approx$16 kpc, the smallest among known absorber-galaxy pairs at $z>5$. ND1 is so heavily dust-obscured that it is detected only at $\lambda\gtrsim3.5\,\mu$m with NIRCam and MIRI. \oiiiemi$\lambda$5007, H$\alpha$, and \ciiemi\ 158 $\mu$m emission anchor the systemic redshift at $z_{\rm sys}=7.0321\pm0.0007$, revealing a $\sim$50 km s$^{-1}$ blueshift of the absorbing gas relative to the host that suggests outflow-driven metal transport into the halo. The implied outflow rate ($\dot{M}_{\rm out}\approx64\,M_\odot$ yr$^{-1}$) and mass-loading factor ($\eta\approx3$) indicate an energetic, metal-enriched wind launched from a dust-obscured starburst. While the ionization state and [Mg/Fe] ratio follow the empirical cosmic trends, the absorber shows an enhanced carbon-to-oxygen ratio, consistent with the yields of Population III stellar explosions and with the host's star-formation history extending to $z\gtrsim8$, potentially a rare fossil record of primordial metal enrichment. Our findings show that intense star-formation feedback can enrich the CGM/IGM well into the Epoch of Reionization, and the NIR-dark host implies that rest-UV surveys may systematically miss the dusty sources that drive early metal enrichment.
\end{abstract}


\section{Introduction}
\label{sec1: intro}

The circumgalactic medium (CGM), the diffuse, multiphase gaseous envelope permeating the dark matter halos of galaxies out to their virial radii, plays an important role in regulating galaxy formation and evolution \citep{Tumlinson+2017_CGM, Faucher-Giguere+2023_CGM}.
Acting as the interface between the interstellar medium in the host galaxy and the large-scale pristine intergalactic medium (IGM), the CGM mediates the inflows and outflows of gas and metals that govern star formation, stellar feedback, and the long-term baryon cycle.
However, its intrinsically low surface brightness renders direct emission mapping exceedingly difficult.
Extended Lyman-$\alpha$ emission has now been shown to be common around high-redshift quasars \citep{Farina+2017_Lya, Cai+2018_Lyn, Arrigoni-Battaia+2019_Lya, Drake+2019_Lya} and galaxies \citep{Wisotzki+2016_Lya, Leclercq+2017_Lya}.
More recently, advances in instrumentation have extended the detection of circumgalactic emission to metal lines \citep{Burchett+2021_MgII_halo, Zabl+2021_MgII_halo, Leclercq+2022_MgII_halo}.
However, at higher redshift ($z\gtrsim 4$), such metal-line CGM emission remains rare: to date, it has been reported only around a handful of the most luminous quasars \citep{Bischetti+2024_CII}, or recovered statistically through stacking analyses \citep{Fujimoto+2019_halo, Bischetti+2019_QSO_CII_halo}.

Intervening metal-line absorbers identified along background quasar sightlines provide a sensitive and broadly applicable probe of circumgalactic gas across cosmic time \citep[e.g., ][]{Becker+2015_abs, Zou+21_abs, Davies+2023_met_abs}.
Because absorption-line detection is essentially independent of the gas surface brightness, it remains effective down to very low column densities that lie well beyond the reach of current emission-line observations, thereby opening a unique window onto the metal-enriched CGM at high redshift.
Spectroscopic surveys have shown such absorption systems to be prevalent out to $z\sim 6.5$ \citep{DOdorico+2023_XQR30, Davies+2023_XQR30}, and their characterization places direct constraints on the chemical enrichment history and physical conditions (e.g., density, temperature, ionization state, and kinematics) of circumgalactic gas.

Current theoretical frameworks attribute early cosmic metal enrichment primarily to star formation and its associated feedback processes \citep[e.g.,][]{Karlsson_REV2013}.
Establishing robust absorber--host galaxy correlations is therefore essential for identifying the sources that are responsible for polluting the CGM and IGM at early epochs.
Cosmological simulations predict that low-ionization metal absorbers are preferentially hosted by galaxies with stellar masses $M_* \sim 10^7$--$10^{9.5}\,M_\odot$, with enriched gas extending to impact parameters of $\sim$150~pkpc \citep{Keating+2016}.
Complementary high-redshift simulations, such as the \textit{Technicolor Dawn} framework, suggest that lower-mass systems ($M_* \sim 10^8\,M_\odot$) may give rise to $z > 5$ absorbers at even larger impact parameters ($\rho$), up to $\sim$300~proper kpc \citep[pkpc;][]{Finlator+2020}.
Empirical determinations of impact parameters, star-formation rates (SFRs), and stellar and halo masses ($M_h$) are therefore critical for discriminating among competing models of metal transport \citep[e.g.,][]{Oppenheimer+2009, Finlator+2013, Hirschmann+2013} and for reconstructing the early chemical enrichment history of the Universe.

Despite considerable observational effort, only $\sim 10$ spectroscopically confirmed absorber--galaxy pairs have so far been identified at $z>5$ \citep{Cai+2017, Diaz+2021, Kashino+2023, Bordoloi+2024_EIGER, Zou+24_ASPIRE}.
This scarcity arises from a combination of physical and observational factors. On the physical side, dark matter halos at $z\gtrsim 5$ typically have masses of $\sim 10^{11}\,M_\odot$ and virial radii of only $\sim 10$--$20$ kpc \citep{Pizzati+2024}, so the associated galaxies are expected to lie within small angular offsets from the background quasar.
On the observational side, rest-frame UV photometric surveys are systematically incomplete for heavily dust-obscured sources, so faint high-$z$ counterparts are easily missed.
Together, these effects make robust identifications of absorber--galaxy pairs at $z>5$ particularly challenging.
Prior to JWST, host galaxy identification relied predominantly on the detection of Ly$\alpha$ \citep[e.g.,][]{Diaz+2014, Diaz+2015, Cai+2017, Sebastian+2026_CGM} or \ciiemi\ emission \citep[e.g.,][]{Kashino+2023, Wu+2021_nat}.
The advent of JWST has substantially expanded these capabilities: \citet{Wu+2023_MgII_abs} recently identified the host galaxy of a $z = 5.43$ \mgii\ absorber through detection of the \oiiiemi\ doublet with NIRCam grism spectroscopy, demonstrating the transformative power of space-based near-infrared spectroscopy for this science case.

The red quasar \tarq\ was first identified in the GOODS-N field by \citet{Fujimoto+2022}, using {\it HST} imaging, and was subsequently confirmed at $z_{\rm qso}=7.19$ through the {\it HST}/WFC3 G141 grism spectrum and NOEMA \ciiemi~158~$\mu$m emission line detection.
Those observations also revealed a dust-obscured companion galaxy in the vicinity of the quasar, at a projected distance of 3\farcs1 ($\sim$16~pkpc at $z=7$).
Subsequent JWST/NIRSpec spectroscopy detected the broad Balmer line from the central engine, yielding a black hole mass of $\log(M_{\rm BH}/M_\odot) = 7.55$ \citep{Fei+2026_GNz7q} and firmly establishing \tarq\ as a super-Eddington accreting quasar in the epoch of reionization.
Detailed inspection of the same NIRSpec spectrum, however, revealed additional absorption features that cannot be attributed to the quasar itself.

In this paper, we report the serendipitous discovery of a metal-rich absorber at $z\approx 7.03$, with metal column densities $\gtrsim1$\,dex above those of typical $z>5$ absorbers, alongside the analysis of a heavily dust-obscured, star-forming galaxy that is associated with this absorber, whose redshift is spectroscopically confirmed through joint JWST/NIRSpec and NOEMA observations along the \tarq\ sightline.
This paper is organized as follows.
Section~\ref{sec2} describes the observations and data reduction.
Section~\ref{sec3} presents the methodology and analytical framework.
Section~\ref{sec4} reports the derived physical properties of the absorbing gas and its host galaxy.
Section~\ref{sec5} discusses the broader astrophysical implications.
Section~\ref{sec6} summarizes our conclusions.
Throughout, we adopt a flat $\Lambda$CDM cosmology with $\Omega_{\rm m}=0.3$, $\Omega_\Lambda=0.7$, and $H_0=70\,\rm km\,s^{-1}\,Mpc^{-1}$, and all magnitudes are reported in the AB system \citep{Oke+1983}.

  \section{Observations and Data Reduction}
 \label{sec2}

\begin{deluxetable*}{llll}
\tabletypesize{\footnotesize}
\tablecaption{Summary of the data used in this work. \label{tab:data}}
\tablewidth{0pt}
\tablehead{
  \colhead{Data} & \colhead{Program (PI)} & \colhead{Depth} & \colhead{Reference}
}
\startdata
NIRSpec G140M/F070LP spectrum of \tarq & GO~4762 (S.~Fujimoto \& G.~Brammer) & 7440\,s & \citealt{Fei+2026_GNz7q} \\
NIRSpec G140M/F100LP spectrum of \tarq & GO~9214 (C.~Mason \& D.~Stark) & 105040\,s & \citealt{Chen+26_SPURS} \\
NIRSpec G235M/F170LP spectrum of \tarq & GO~9214 (C.~Mason \& D.~Stark) & 28448\,s & \citealt{Chen+26_SPURS} \\
\smallskip
\multirow{2}{*}{NIRSpec G395M/F290LP spectrum of \targ} & GO~4762 (S.~Fujimoto \& G.~Brammer) & 7440\,s & \citealt{Fei+2026_GNz7q} \\
 & GO~7935 (F.~Sun \& X.~Lin) & 14881\,s & \citealt{Sun+2025_JWST} \\
\hline
\multirow{3}{*}{\begin{tabular}[c]{@{}l@{}}NIRCam F090W, F115W, F182M,\\ F210M, F356W, F444W\end{tabular}} & GO~1895 (FRESCO; P.~Oesch) & \multirow{3}{*}{\begin{tabular}[c]{@{}l@{}}28.2, 28.4, 28.9,\\ 28.7, 28.1, 27.9\,mag\end{tabular}} & \citealt{Oesch+2023_FRESCO} \\
 & GO~3577 (CONGRESS; E.~Egami) & & \citealt{Egami+2023_CONGRESS} \\
 & GO~4762 (S.~Fujimoto \& G.~Brammer) & & \citealt{Fei+2026_GNz7q} \\
\hline
\multirow{2}{*}{MIRI F1000W, F1280W, F2100W} & GO~4762 (S.~Fujimoto \& G.~Brammer) & \multirow{2}{*}{24.8, 25.2, 23.2\,mag} & \citealt{Fei+2026_GNz7q} \\
 & GO~5407 (MEOW; G.~C.~K.~Leung) & & \citealt{Leung+2024_MEOW} \\
\hline
NOEMA 1\,mm (233\,GHz) continuum & Archival & 23\,$\mu$Jy\,beam$^{-1}$ & \citealt{Fujimoto+2022} \\
NOEMA \ciiemi\ data cube & W25E004 (Q.~Fei \& F.~Walter) & 0.32\,mJy\,beam$^{-1}$ & This work \\
\enddata
\tablecomments{
    For the spectra, the depth is the total on-source integration time. 
    For the imaging, it is the $5\sigma$ depth (AB) at the position of \targ\ in each listed filter. 
    The NIRCam and MIRI data of all programs are combined into one mosaic per filter. 
    For the 1\,mm continuum, it is the $1\sigma$ rms noise of the dust continuum map, and for the \ciiemi\ data cube, it is the $1\sigma$ rms noise per 120\,km\,s$^{-1}$ channel. 
    In the NIRCam imaging, F090W, F115W, and F356W come from GO~3577, F182M and F210M from GO~1895 and GO~4762, and F444W from GO~1895 and GO~4762. 
    In the MIRI imaging, F1280W comes from GO~4762 and F1000W and F2100W from GO~5407.}
\end{deluxetable*}

\begin{figure*}
    \centering
    \includegraphics[width=\linewidth]{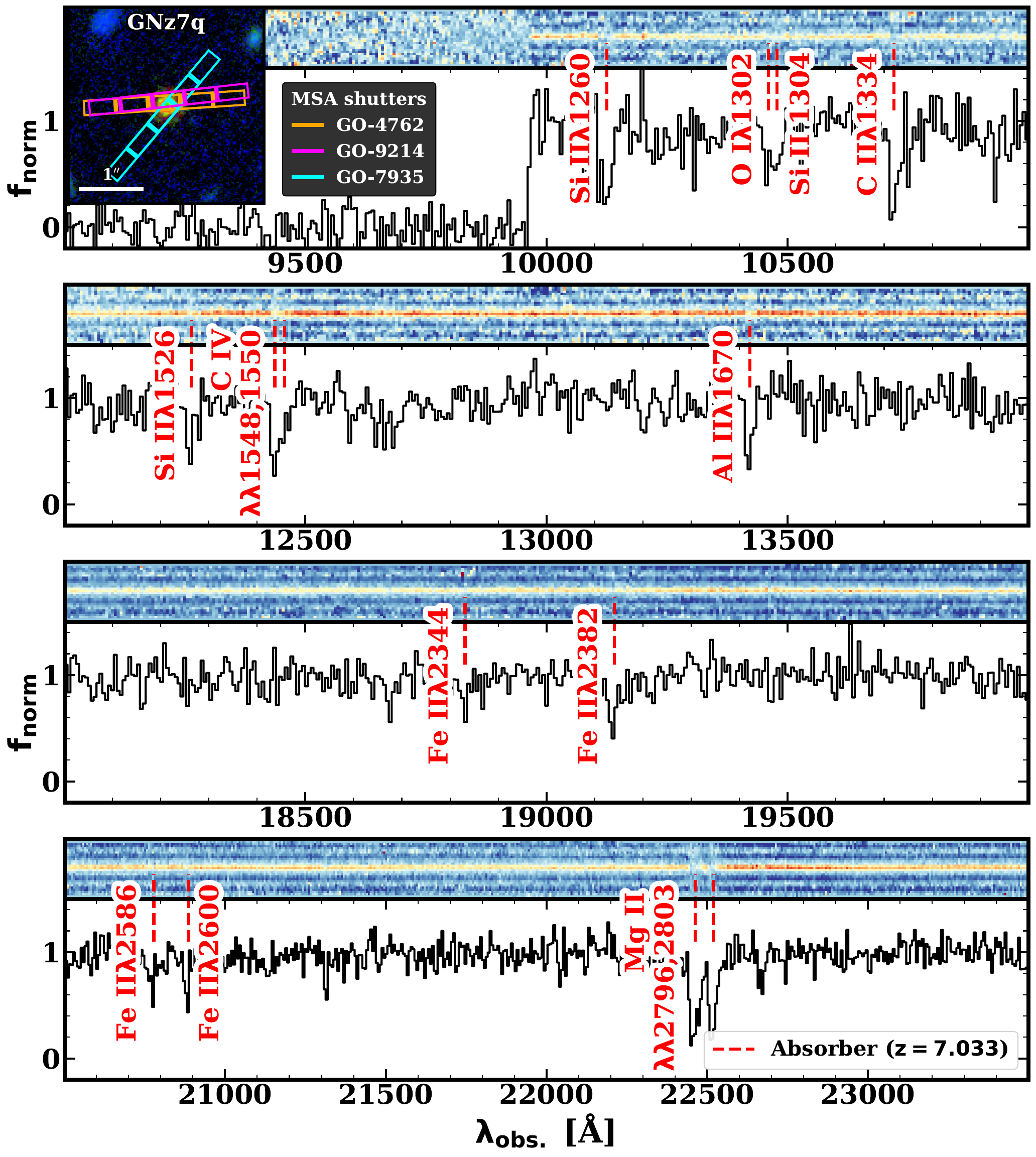}
    \caption{
    {The original 2D and normalized 1D rest-frame UV spectrum of \tarq\ (G140M/F070LP, GO~4762).}
    The spectra are shown as normalized flux versus observed wavelength ($\lambda_{\rm obs}$). Each panel presents a two-dimensional (2D) spectral trace at the top and the corresponding extracted, one-dimensional (1D) spectrum below it. These spectra reveal 16 metal absorption transitions from nine ions spanning neutral to highly ionized species (Table~\ref{tab:absorption}), which are identified and annotated with red dashed vertical lines and corresponding text labels (legend in the bottom panel).
    The inset in the top-left corner of the first row shows a NIRCam false-color image (F444W/F210M/F090W) of a $3\farcs2\times3\farcs2$ field centered on \tarq, overlaid with the NIRSpec MSA shutters of the three spectroscopic programs (orange: GO~4762; magenta: GO~9214; cyan: GO~7935; see the legend to the right of the inset); the scale bar is $1''$ ($\approx5.2$\,pkpc at $z=7.03$).
    }
    \label{fig01: norm spec}
\end{figure*}

This study uses all available JWST and NOEMA data targeting this field, including our own PI program and archival treasury programs.
Table~\ref{tab:data} summarizes the programs used in this work, and the inset panel of Figure~\ref{fig01: norm spec} shows the NIRSpec MSA shutters of the three spectroscopic programs overlaid on the false-color NIRCam image of the background quasar \tarq.
The following sections describe the data reduction and analysis of these data.

 \subsection{NIRSpec}
\label{sec2.1}

The deep spectroscopic follow-up of \targ, a dust-obscured, star-forming galaxy at $z=7.033$ located $\sim$3\farcs1 to the North--East of the background red quasar \tarq, and the host of the metal absorption system studied in this work, was obtained with JWST/NIRSpec as part of GO~4762 (PIs: S.~Fujimoto \& G.~Brammer), an approved Cycle-2 program designed to obtain deep NIRSpec G140M/F070LP and G395M/F290LP spectra ($R\sim 1000$) of the super-Eddington red quasar \tarq\ at $z=7.2$ and to characterize the properties of nearby galaxies through multi-object spectroscopy \citep{Fei+2026_GNz7q}.
Each grating was observed with a total on-source integration time of 7440\,s ($\simeq2.1$\,hr) using a three-shutter nod pattern.
The MSA configuration of GO~4762 was designed with the target-prioritization and mask-optimization algorithm developed for the UNCOVER and RUBIES surveys.
We refer to \citet{Price+2025} and \citet{deGraaff+2025_RUBIES} for a full description and summarize here only its application to our program.
Candidate targets were ranked in two tiers: (1) individual sources of particular scientific interest, placed first, and (2) statistical samples of galaxies of broader interest (e.g., photometric $z>4$ galaxies) , with equal priority within each catalog.
The optimization assigned shutters to 241 primary sources, including \tarq \ (at the center of the configuration, to secure its full integration) and \targ.
\targ\ was placed in the first tier because of its 1-mm continuum detection in previous NOEMA observations \citep{Fujimoto+2022}.

We additionally employ NIRSpec G140M-grating and G235M-grating data from the SPURS program (PID~9214; PIs: C.~Mason \& D.~Stark), which was designed to characterize rest-frame UV properties in a general high-redshift galaxy population \citep{Mason+2025_SPURS} (more details about the target selection, the observation setup, and results are given in \citealt{Chen+26_SPURS}).
The red quasar \tarq\ was also included in the MSA mask design of that program,
with total on-source integration times of $\sim$105~ks ($\simeq$29.2~hr) in G140M-F100LP and $\sim$28.4~ks ($\simeq$7.9~hr) in G235M-F170LP.
We also employ NIRSpec G395M-grating data from GO~7935 (PIs: F.~Sun \& X.~Lin), a Cycle-3 program that measures the emergence rate of AGN in the legacy field \citep{Sun+2025_JWST}.
The red quasar \tarq\ and the nearby companion \targ\ were also included in the GO~7935 mask design.

The NIRSpec data from all three programs (GO~4762, GO~9214, and GO~7935) were processed with the same \texttt{msaexp} pipeline \citep{Brammer2023} and the same procedure (from the uncalibrated exposures to the optimally extracted 1D spectra, including background subtraction from the nodded exposures and the point-source path-loss correction), following the standard practice adopted in previous JWST/NIRSpec MOS work \citep[e.g.,][]{Heintz+2025_PRIMAL, deGraaff+2025_RUBIES, Naidu+2025_BH*, Valentino+2025_outflow}. Each spectrum is reduced, extracted, and analyzed independently, and the rest-frame UV spectra are not co-added.

\subsection{NIRCam \& MIRI images}
\label{sec2.2: NIRCam}

The NIRCam imaging data are retrieved from several programs covering the GOODS-N field, including GO~4762 (PIs: S.~Fujimoto \& G.~Brammer) and the FRESCO survey (GO~1895; \citealt{Oesch+2023_FRESCO}). The programs contributing to each filter at the position of \targ\ are summarized in Table~\ref{tab:data}.
The source was observed with NIRCam through the F090W, F115W, F182M, F210M, F356W, and F444W filters.
Our reduction follows the standard procedure used for JWST/NIRCam analyses of the GOODS-N field \citep[e.g., ][]{Oesch+2023_FRESCO, Eisenstein+2023_JADES, Atek+2025_glimpse}.
The processing of NIRCam imaging data commenced with the uncalibrated raw files, using the JWST Science Calibration Pipeline (v12.0.9) and CRDS context file \texttt{jwst\_1321.pmap}.
Beyond the standard STScI procedures, we implemented custom steps to mitigate artifacts, including the removal of so-called ``snowballs'' (large, extended cosmic-ray strikes), wisps, cosmic rays, persistence signals from earlier exposures, and diffraction spikes, alongside an initial {1/\emph{f}}-noise correction.
Following the Stage 2 pipeline processing, we applied a secondary correction for residual {1/\emph{f}} noise via sigma-clipping and performed two-dimensional background subtraction using the \texttt{sep} package.
The data were then co-added through Stage 3 of the pipeline to produce the final astrometrically aligned mosaics (aligned to Gaia astrometry of stars in the field of view) with pixel scales of 20 mas~pix$^{-1}$ for short-wavelength (SW) bands and 40 mas~pix$^{-1}$ for long-wavelength (LW) bands.

The MIRI imaging data are also retrieved from two programs: F1280W from GO~4762 (PIs: S.~Fujimoto \& G.~Brammer) and F1000W and F2100W from the MEOW survey (GO~5407; \citealt{Leung+2024_MEOW}).
The F1280W data were reduced following the standard procedures established by the Dawn JWST Archive (DJA; \citealt{Valentino+2023_QGs}) and widely used in public surveys \citep[e.g., PRIMER;][]{Dunlop+2021_PRIMER}. Specifically, we retrieved the Level-2 products and processed them using the \texttt{grizli} software package \citep{Brammer+2021_grizli, Brammer+2022_grizli}, adopting the methodology described by \citet{Valentino+2023_QGs}.

To utilize the F1000W and F2100W bands, the data publicly available in September 2025 for the GOODS-N field, jointly with the MEOW data (PID~5407, PI: G.C.K.~Leung; \citealt{Leung+2024_MEOW}), were reduced with the JWST Rainbow pipeline following the methodology described in \cite{Perez-Gonzalez+2024a_MIRI}.
The observations came from PIDs 1181, 1264, 2926, 4762, and 5407, and mosaics were constructed for the F560W, F770W, F1000W, F1280W, F1800W, and F2100W bandpasses.
\tarq\ and \targ\ are covered only in the 10~$\mu$m, 12.8~$\mu$m, and 21~$\mu$m bandpasses.
Details about this reduction (jointly with a similar effort in other observed fields such as GOODS-S, EGS, or COSMOS) will be presented by P\'erez-Gonz\'alez et al. (in prep.). We briefly describe the methodology here.
The JWST Rainbow pipeline is built on top of the official STScI JWST calibration pipeline.
It applies additional custom steps that are designed to remove residual, spatially varying background structure from the MIRI images (a step we hereafter refer to as ``background homogenization'').
For this paper, the reduction was carried out with JWST pipeline version v1.19.41 and CRDS reference-file map \texttt{jwst\_1413.pmap} (which defines the calibration reference files applied at each pipeline stage).
A key component of this workflow is the ``superbackground'' strategy, in which background maps for individual exposures are constructed using the full set of campaign images, with known discrete sources masked so that only the diffuse background is used to build the map.
This approach yields a highly homogeneous background in terms of both level and noise.
Quantitatively, this method reduces the standard deviation of the diffuse background in individual Stage-2 calibrated exposures (the \texttt{*\_cal.fits} files produced by the pipeline) by a factor of $\sim$3.5 relative to the standard STScI JWST calibration pipeline v1.19.41, resulting in a sensitivity improvement of 0.8 mag in the final mosaic. Further details are provided in Appendix~A of \cite{Perez-Gonzalez+2024a_MIRI}, and by \cite{2025A&A...696A..57O} and \cite{2024ApJ...976..224A}, who report similar improvements in the noise of the final mosaics.

\subsection{NOEMA observation}
\label{sec2.3:noema}
This field had previously been targeted with NOEMA by \citet{Fujimoto+2022}, whose Band-3 observations delivered the original 1\,mm dust-continuum map and the \ciiemi\ detection of the background quasar \tarq. Those data also revealed a bright dust-continuum source coincident with \targ, but the frequency setup was tuned to the redshifted \ciiemi\ line of \tarq\ and did not cover \ciiemi\ at the redshift of \targ. A dedicated follow-up was therefore required to secure the \ciiemi\ line of \targ.

We carried out new Band~3 (1\,mm) observations targeting the \ciiemi\ transition using the Institut de Radio astronomie Millim\'etrique (IRAM) NOrthern Extended Millimeter Array (NOEMA) under the project code W25E004 (PIs: Q.~Fei \& F.~Walter).
The phase center was aligned with the \targ\ coordinates derived from the JWST/NIRCam F444W imaging.
The {\it PolyFix} correlator was employed in two spectral setups, providing simultaneous coverage of two 7.744\,GHz-wide sidebands.
The upper sideband (USB) was centered at 236.591\,GHz to capture the redshifted \ciiemi\ emission, while the lower sideband (LSB) was dedicated to measuring the dust continuum.
Observations were executed between 2025 December 13 and 18 in the extended C configuration, yielding a total integration time of 20 hours.
Data reduction was performed remotely at IRAM Grenoble utilizing the {\it CLIC} package within the GILDAS software framework.
The resulting data achieve a root-mean-square (rms) noise level of 0.32$\rm mJy\,beam^{-1}$ per 120$\rm km\,s^{-1}$ channel, with a synthesized beam full width at half maximum (FWHM) of 1\farcs22 $\times$ 1\farcs08.
The dust continuum of \targ\ is also recovered in the LSB of these new observations (at $\approx221$\,GHz), at a level and morphology consistent with the archival map of \citet{Fujimoto+2022}.
In this work, the archival continuum data are used only for the 1\,mm photometric point of \targ\ in the SED fit (Section~\ref{host}).
We adopt the archival 232\,GHz measurement for the SED fit because the new LSB continuum is consistent with it within the uncertainties and therefore does not change the fit, while a joint reduction of both epochs into a deeper continuum map is beyond the scope of this paper. All \ciiemi\ measurements of \targ\ are based exclusively on the new W25E004 data.

\begin{figure*}[h]
    \centering
    \includegraphics[width=0.99\linewidth]{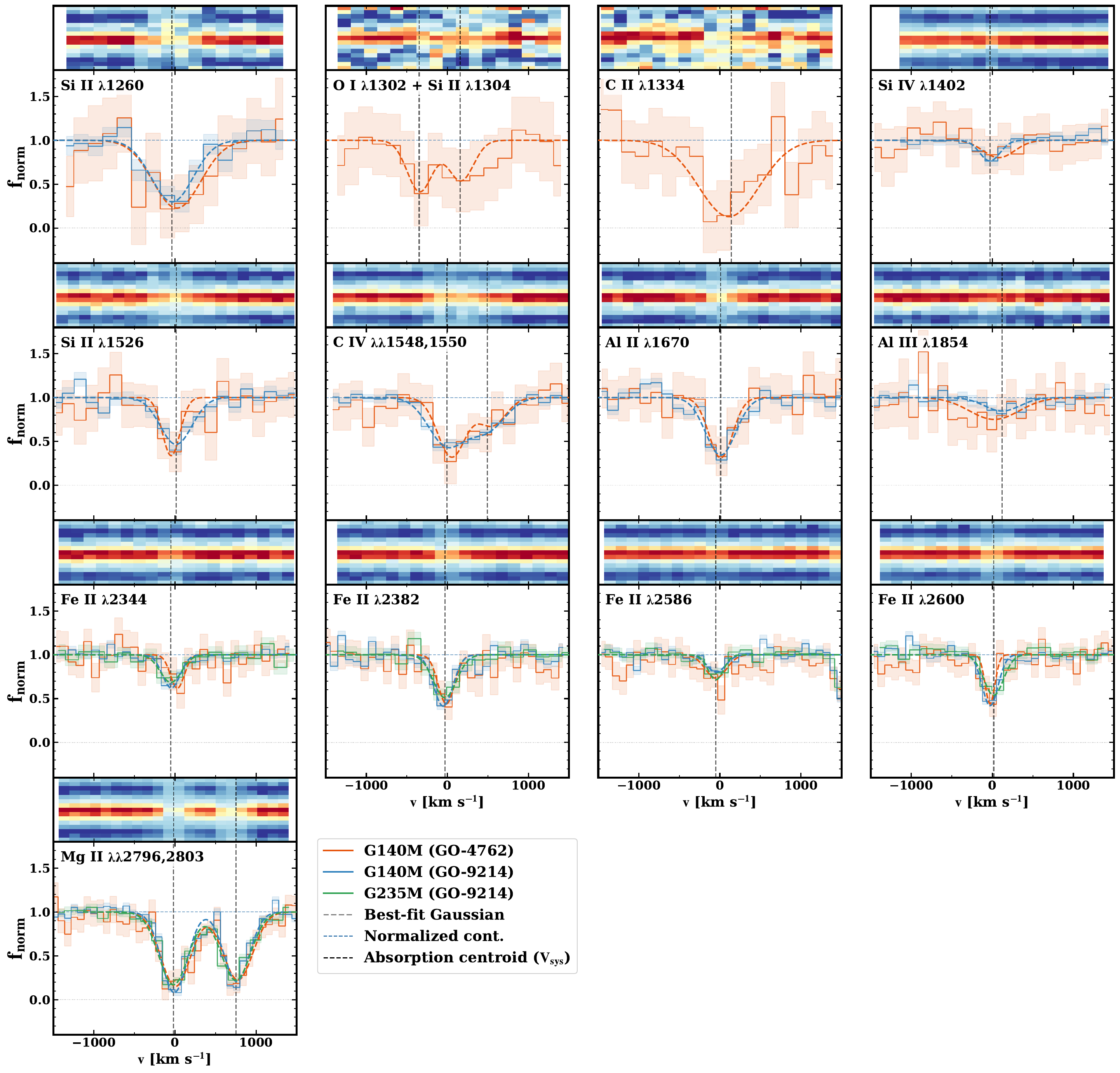}
    \caption{
    {Zoom-in view of all detected absorption lines with all three NIRSpec spectra overlaid.}
    Each panel shows the 2D spectral stamp (top) and the 1D normalized flux from all available gratings (bottom): G140M-F070LP from GO~4762 (orange), G140M-F100LP from GO~9214 (blue), and G235M-F170LP from GO~9214 (green).
    Dashed curves with the same color show the corresponding best-fit Gaussian model.
    The black dashed vertical line marks the centroid of each absorption line, and the velocity zero point corresponds to the mean redshift of the absorption lines ($z_{\rm abs}=7.0316$).
    Panels with only a single spectrum shown indicate that the other two spectra do not cover that wavelength.
    }
    \label{fig02:EW}
\end{figure*}

Image synthesis and deconvolution were conducted using the MAPPING routine within the GILDAS environment.
Dirty maps were generated from the calibrated visibilities applying natural weighting without $uv$-tapering, a choice motivated by our need to maximize point-source sensitivity to the moderately extended \ciiemi\ emission of \targ\ rather than to resolve its spatial structure.
Primary beam correction was not applied; given the source's location at the phase center, the attenuation at the edges of our extraction apertures is negligible and does not affect the subsequent analysis.
The maps were cleaned down to a threshold of 2$\sigma$ (where $\sigma$ represents the local rms noise in the dirty map) within a circular mask of radius $r=5''$.

We generated a dirty data cube with a spectral resolution of 120\,$\rm km\,s^{-1}$.
The underlying dust continuum was estimated from line-free channels via a first-order polynomial fit across the $\sim$7.6\,GHz sidebands using the GILDAS {\it UV\_BASELINE} task.
This continuum model was subsequently subtracted directly in the $uv$-plane to produce the final line-only data cubes.
Finally, we constructed a velocity-integrated intensity (moment-0) map of the \ciiemi\ transition by integrating the continuum-subtracted cube over a velocity range of $[-300, +300]\,\rm km\,s^{-1}$ centered on the expected redshifted line frequency, an interval that is wide enough to include the full line profile but narrow enough to minimize the noise from adjacent line-free channels. The resulting spatial intensity distribution and the extracted 1D spectrum are presented in Section~\ref{host}.

\section{Data Analysis}
\label{sec3}

In this section we describe how the different data sets introduced in Section~\ref{sec2} are combined into a coherent physical picture of the system. Section~\ref{sec3.1} presents the identification and equivalent-width measurement of the metal absorption lines in the NIRSpec spectra of the background quasar \tarq, and derives their column densities from a multi-ion curve-of-growth analysis. Section~\ref{host} then combines the JWST/NIRSpec, JWST/NIRCam, JWST/MIRI, and NOEMA data of \targ\ to constrain the systemic redshift, stellar mass, star-formation rate, and dust extinction of the absorber's host galaxy.

\subsection{Metal Absorption Lines}
\label{sec3.1}

\begin{deluxetable*}{lcccccc}
\tabletypesize{\footnotesize}
\tablecaption{Absorption Line Properties \label{tab:absorption}}
\tablewidth{0pt}
\tablehead{
  \colhead{Species} & \colhead{$\lambda_\mathrm{rest}$} &
  \colhead{$W_0$ (G140M)} & \colhead{$W_0$ (G140M)} & \colhead{$W_0$ (G235M)} &
  \colhead{$\log_{10}(N / \mathrm{cm^{-2}})$} & \colhead{$|v_{\rm out}|$} \\
  \colhead{} & \colhead{(\AA)} & \colhead{(\AA)} & \colhead{(\AA)} & \colhead{(\AA)} & \colhead{} & \colhead{(km\,s$^{-1}$)} \\
  \colhead{(1)} & \colhead{(2)} & \colhead{(3)} & \colhead{(4)} & \colhead{(5)} & \colhead{(6)} & \colhead{(7)}
}
\startdata
  SiII\,$\lambda$1260 & 1260.42 & $4.161 \pm 2.610$ & $1.976 \pm 0.274$ & \nodata$^\dagger$ & $14.795^{+0.067}_{-0.077}$ & $571 \pm 101$ \\
  OI\,$\lambda$1302 & 1302.17 & $0.850 \pm 0.274$ & \nodata & \nodata & $15.164^{+0.150}_{-0.223}$ & $347 \pm 138$ \\
  SiII\,$\lambda$1304 & 1304.37 & $0.925 \pm 0.274$ & \nodata & \nodata & $14.795^{+0.067}_{-0.077}$ & $347 \pm 138$ \\
  CII\,$\lambda$1334 & 1334.53 & $2.591 \pm 0.441$ & \nodata & \nodata & $15.641^{+0.288}_{-0.269}$ & $545 \pm 125$ \\
  {SiIV\,$\lambda$1402} & {1402.77} & {$1.842 \pm 1.668$} & {$0.343 \pm 0.093$} & \nodata & {$13.921^{+0.105}_{-0.159}$} & $237 \pm 192$ \\
  SiII\,$\lambda$1526 & 1526.71 & $2.499 \pm 2.003$ & $1.341 \pm 0.150$ & \nodata & $14.795^{+0.067}_{-0.077}$ & $456 \pm 75$ \\
  CIV\,$\lambda$1548 & 1548.20 & $1.336 \pm 0.334$ & $1.536 \pm 0.117$ & \nodata & $>14.585$ & $376 \pm 102$ \\
  CIV\,$\lambda$1550 & 1550.77 & $0.492 \pm 0.288$ & $1.105 \pm 0.106$ & \nodata & $>14.585$ & $376 \pm 102$ \\
  AlII\,$\lambda$1670 & 1670.79 & $2.115 \pm 1.374$ & $1.748 \pm 0.126$ & \nodata & $13.791^{+0.045}_{-0.054}$ & $303 \pm 58$ \\
  {AlIII\,$\lambda$1854} & {1854.72} & {$1.767 \pm 1.363$} & {$0.904 \pm 0.480$} & {\nodata} & {$13.796^{+0.214}_{-0.408}$} & $678 \pm 220$ \\
  FeII\,$\lambda$2344 & 2344.21 & $1.567 \pm 0.928$ & $0.821 \pm 0.096$ & $0.902 \pm 0.245$ & $14.043^{+0.038}_{-0.045}$ & $68 \pm 35$$^\ddagger$ \\
  FeII\,$\lambda$2382 & 2382.76 & $1.503 \pm 0.676$ & $1.405 \pm 0.117$ & $1.403 \pm 0.193$ & $14.043^{+0.038}_{-0.045}$ & $223 \pm 65$ \\
  FeII\,$\lambda$2586 & 2586.65 & $1.024 \pm 0.691$ & $0.543 \pm 0.122$ & $0.658 \pm 0.165$ & $14.043^{+0.038}_{-0.045}$ & $70 \pm 156$$^\ddagger$ \\
  FeII\,$\lambda$2600 & 2600.17 & $0.752 \pm 0.270$ & $1.261 \pm 0.073$ & $1.248 \pm 0.130$ & $14.043^{+0.038}_{-0.045}$ & $47 \pm 23$$^\ddagger$ \\
  MgII\,$\lambda$2796 & 2796.35 & $3.579 \pm 0.340$ & $3.329 \pm 0.084$ & $3.415 \pm 0.110$ & $14.277^{+0.019}_{-0.021}$ & $258 \pm 25$ \\
  MgII\,$\lambda$2803 & 2803.53 & $3.323 \pm 0.321$ & $2.954 \pm 0.082$ & $3.357 \pm 0.106$ & $14.277^{+0.019}_{-0.021}$ & $258 \pm 25$ \\
\enddata
\tablecomments{%
  Columns: (1) absorbing species and rest-frame wavelength of the transition; (2) rest-frame vacuum wavelength of the transition in \AA; (3)--(5) rest-frame equivalent width, $W_0 \equiv \int\!\left[1 - f_{\rm norm}(\lambda)\right]{\rm d}\lambda$, measured independently from the three NIRSpec spectra used in this work: G140M from GO~4762 (col. 3), G140M from the SPURS program GO~9214 (col. 4), and G235M from GO~9214 (col. 5); (6) column densities derived from a multi-ion single-component curve of growth (COG) analysis at the fiducial Doppler parameter $b=200\,\mathrm{km\,s^{-1}}$ (see Section~\ref{sec3.2}); (7) outflow velocity computed following \citet{Fiore+2017_AGN_of} as $|v_{\rm out}| = |\Delta v| + 2\sigma_{\rm int}$, where $\Delta v$ is the fitted centroid offset from the systemic redshift $z_{\rm sys}=7.0321$ and $\sigma_{\rm int}$ is the LSF-deconvolved intrinsic velocity dispersion (adopting $\sigma_{\rm inst}\approx106\,\mathrm{km\,s^{-1}}$ for the NIRSpec MSA).
  All $W_0$ values are derived from fitting rather than from direct spectrum integration. They are the integrals of the best-fit local Gaussian models (Section~\ref{sec3.1}), with $1\sigma$ uncertainties obtained from MCMC posterior sampling (median $\pm$ half the 16th--84th percentile range).
  For each transition, the spectrum with the smallest measurement uncertainty is adopted as the primary input to the COG analysis.\\
  $^\dagger$ \nodata\ indicates that the transition is not covered by that spectrum.\\
  $^\ddagger$ Intrinsic line width unresolved after deconvolution with the line spread function. $|v_{\rm out}|$ then reduces to $|\Delta v|$ and should be interpreted as a lower limit.
}
\end{deluxetable*}

This subsection focuses on the JWST/NIRSpec spectra of the background quasar \tarq\ and, in particular, on the rest-frame ultraviolet wavelength window ($\lambda_{\rm rest}\approx 1200$--$3000$\,\AA) that contains the low- and intermediate-ionization metal transitions used to identify the CGM absorber and to derive its equivalent widths and column densities.

 We model the continuum of the JWST/NIRSpec spectrum of \tarq\ adopting the methodology established for high-$z$ quasar studies \citep{Christensen+2023_qso_abs}.
We use the \texttt{Astrocook} software \citep{Cupani+2022_Astrocook} to generate a global continuum fit, using a set of user-defined continuum anchor nodes distributed at approximately equal velocity intervals of 1000\,$\rm km\,s^{-1}$, together with an iterative $5\sigma$ rejection that excludes outlier pixels associated with narrow absorption features.
A cubic spline is subsequently interpolated between the remaining nodes; it is smooth enough to follow the broad emission-line wings of the quasar without introducing high-frequency oscillations that could bias the absorption-line measurements.
The 1000\,$\rm km\,s^{-1}$ spacing is dense enough to trace the broad emission-line profiles of the quasar (typical FWHM $\sim 3000$\,$\rm km\,s^{-1}$) with several nodes per emission line; nodes that happen to fall on top of a strong emission line are automatically deprioritized by the $5\sigma$-rejection step, so that the interpolated continuum does not carve a spurious dip into the emission profile.
Before normalization, we visually inspect the spectrum in the interactive \texttt{Astrocook} graphical interface and manually reposition individual nodes that are visibly biased by narrow absorption or noise spikes, ensuring that the local continuum follows the smooth envelope of the emission-line profile.
The resulting normalized flux, $f_{\rm norm}(\lambda)$, and corresponding error spectra, $\sigma(f_{\rm norm,\lambda})$, are then computed by dividing the observed spectrum by the fitted continuum.
Figure~\ref{fig01: norm spec} presents the normalized spectra overlaid with the identified absorption systems marked in red.

We identified and classified the metal absorption features through visual inspection of the observed spectrum, focusing on the ensemble of low- and intermediate-ionization transitions that are routinely detected toward bright quasars \citep[e.g., ][]{Chen+2010_MgII, Werk+2014_COS, Simcoe+2011_CIV, Cooper+2019_CGM}.
These transitions have been widely employed as diagnostics of the physical and chemical state of the circumgalactic medium at both low and high redshift, notably in surveys of quasar damped Lyman-$\alpha$ absorbers \citep[DLAs; e.g.,][]{Prochaska+2003_qso} and of gamma-ray-burst afterglow absorption systems \citep{Prochaska+2006_GRB, Christensen+2011_GRB}, making them a well-established baseline for our line identification.
Absorption systems were identified through the strongest UV doublets, \ion{Mg}{2}\ $\lambda\lambda 2796,2803$ and \ion{C}{4}$\lambda\lambda 1548,1550$, whose precise wavelength separations and oscillator-strength-fixed intensity ratios anchor an unambiguous system redshift.
Additional single transitions were accepted as members of a given system only when they satisfied two joint criteria: (i) a significance of at least $3\sigma$ above the local continuum, and (ii) a wavelength centroid consistent, within the measurement uncertainty, with the redshift set by the doublet.
Applying this procedure to the observed spectrum, we identify one high-confidence absorber at $z = 7.0315\pm0.0014$, whose associated transitions span a broad range of ionization stages, from neutral and singly ionized species (e.g., \oi, \ion{C}{2}, \ion{Mg}{2}) to intermediate- and high-ionization lines (e.g., \ion{C}{4}), providing a consistent multi-species diagnostic of the underlying absorbing gas.

\begin{figure*}
    \centering
    \includegraphics[width=\linewidth]{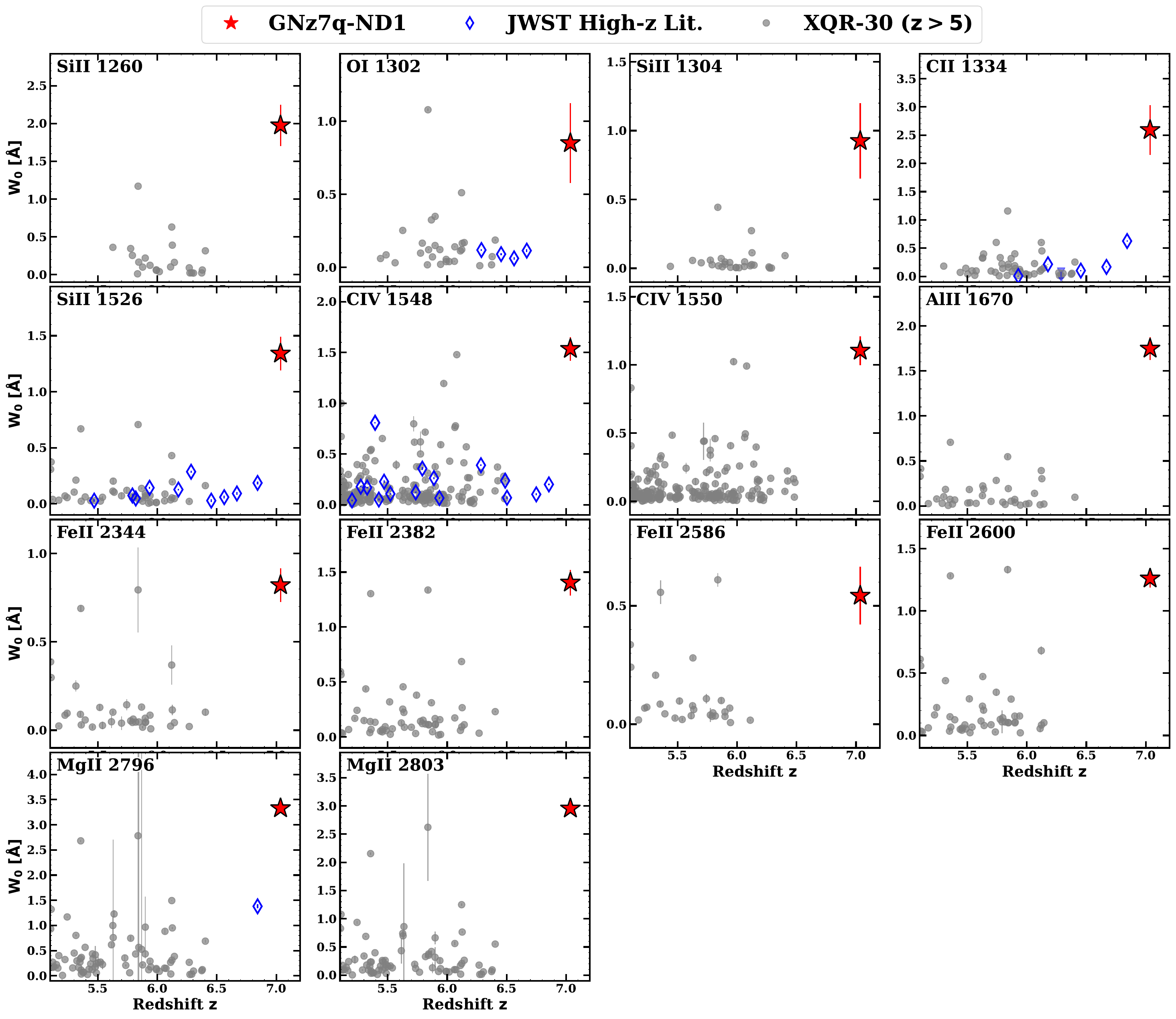}
    \caption{
    {Rest-frame equivalent widths ($W_0$) of metal absorption lines as a function of redshift.} The measurement for our target is denoted by the red star.
    For comparison, samples from the literature are represented by gray circles \citep{Davies+2023_met_abs} and blue diamonds \citep{Christensen+2023_qso_abs}.
    Species names are labeled in the upper-left corner of each panel.
    }
    \label{fig08:EW}
\end{figure*}

We estimate the equivalent width ($W_0$) of the absorption lines using the standard definition in wavelength units:
\begin{align}
    W_0 \equiv \int \left(1 - e^{-\tau_\lambda}\right) d\lambda
             = \lambda_0 \int \frac{d\nu}{\nu_0}\left(1 - e^{-\tau_\nu}\right),
    \label{eq1: EW}
\end{align}
where $\lambda_0$ ($\nu_0$) denotes the rest-frame central wavelength (frequency) of the transition, and $\tau_\lambda$ ($\tau_\nu$) represents the optical depth as a function of wavelength (frequency).
All $W_0$ values reported in Table~\ref{tab:absorption} are in units of \AA.
Empirically, this quantity can be approximated by integrating the normalized spectral profile: ${\rm EW} \approx \int [1 - f_{\rm norm}(\lambda)]\,d\lambda$, an operation that is formally independent of spectral resolution for isolated transitions.
In our data, however, the moderate G140M resolution causes several closely spaced transitions to blend (e.g., \oi\ and \siii$\lambda$1304), so a simple direct integration cannot cleanly separate their contributions.
To disentangle blended features and to propagate the measurement errors self-consistently, we adopted a parametric modeling approach.
We performed Gaussian fitting to the normalized spectra within isolated windows surrounding each detection, deriving the EWs and associated uncertainties directly from the posterior distribution of the best-fit model.
For complex, heavily blended systems where unique deconvolution is required, we employed a multi-component double-Gaussian decomposition.
In these instances, we fixed the relative line centroids to their theoretical wavelength separations, while the line widths and amplitudes of each component were treated as free parameters (i.e., the line widths were not fixed) to maximize the flexibility and accuracy of the fit.
For the purposes of $W_0$ measurement, Gaussian fitting provides a robust estimate.
The strengths of these absorption lines are characterized in detail via the curve-of-growth analysis in the following section.
The best-fit results and the EWs are listed in Table~\ref{tab:absorption}, and the corresponding zoom-in views of every detected transition are shown in Figure~\ref{fig02:EW}.

The three spectra were obtained independently with different MSA configurations, so \tarq\ is located at different positions within the $0\farcs20\times0\farcs46$ shutters (inset of Figure~\ref{fig01: norm spec}).
However, since the background source is a point-like quasar that is fully contained in each shutter, all three spectra probe the same pencil-beam sightline through the absorber, with a transverse extent set by the quasar continuum region rather than by the shutter size (which corresponds to only $\approx1\times2.4$\,pkpc at $z=7.03$).
Differences in the shutter position mainly change the slit losses, which scale the continuum and the absorbed flux by the same factor.
Therefore, the normalized spectrum, and hence the equivalent widths of the absorption lines, are unaffected.
Consistent with this, the $W_0$ of transitions covered by more than one spectrum agree within $\lesssim2\sigma$ (Table~\ref{tab:absorption}; e.g., $W_0=3.58\pm0.34$, $3.33\pm0.08$, and $3.42\pm0.11$\,\AA\ for \ion{Mg}{2}\,$\lambda2796$).
The larger scatter of some weak or blended lines in the GO~4762 spectrum (e.g., \ion{Si}{2}\,$\lambda1260$) is fully accounted for by their measurement uncertainties (those lines are not covered by any other spectrum due to the detector gap).

 We compare the measured EWs for each ionic transition in \targ\ against literature samples of high-redshift metal absorbers \citep{Davies+2023_met_abs, Christensen+2023_qso_abs}.
As demonstrated in Figure~\ref{fig08:EW}, our target exhibits the largest EWs observed to date at $z>6$: $W_0=3.33\pm0.08$\,\AA\ for \ion{Mg}{2}\,$\lambda2796$ and $W_0=2.59\pm0.44$\,\AA\ for \ion{C}{2}\,$\lambda1334$ exceed the largest values among the $z>6$ absorbers of \citet{Davies+2023_met_abs} and \citet{Christensen+2023_qso_abs} by factors of $\approx2.2$ and $\approx4.1$, respectively, and those of the \ion{Si}{2}, \ion{Al}{2}, and \ion{Fe}{2} transitions by factors of $\approx2$--4.5.
At $5<z<6$, only the strongest absorber of \citet[][at $z=5.84$]{Davies+2023_met_abs} has comparable or larger $W_0$ in \ion{O}{1}, \ion{C}{4}, and \ion{Fe}{2}.
Since $W_0$ increases monotonically with column density (Section~\ref{sec3.2}), these strong absorption features indicate that we are probing a dense, metal-enriched CGM in the early Universe.

\subsection{Curve of Growth}
 \label{sec3.2}

\begin{figure}[h]
    \centering
    \includegraphics[width=\linewidth]{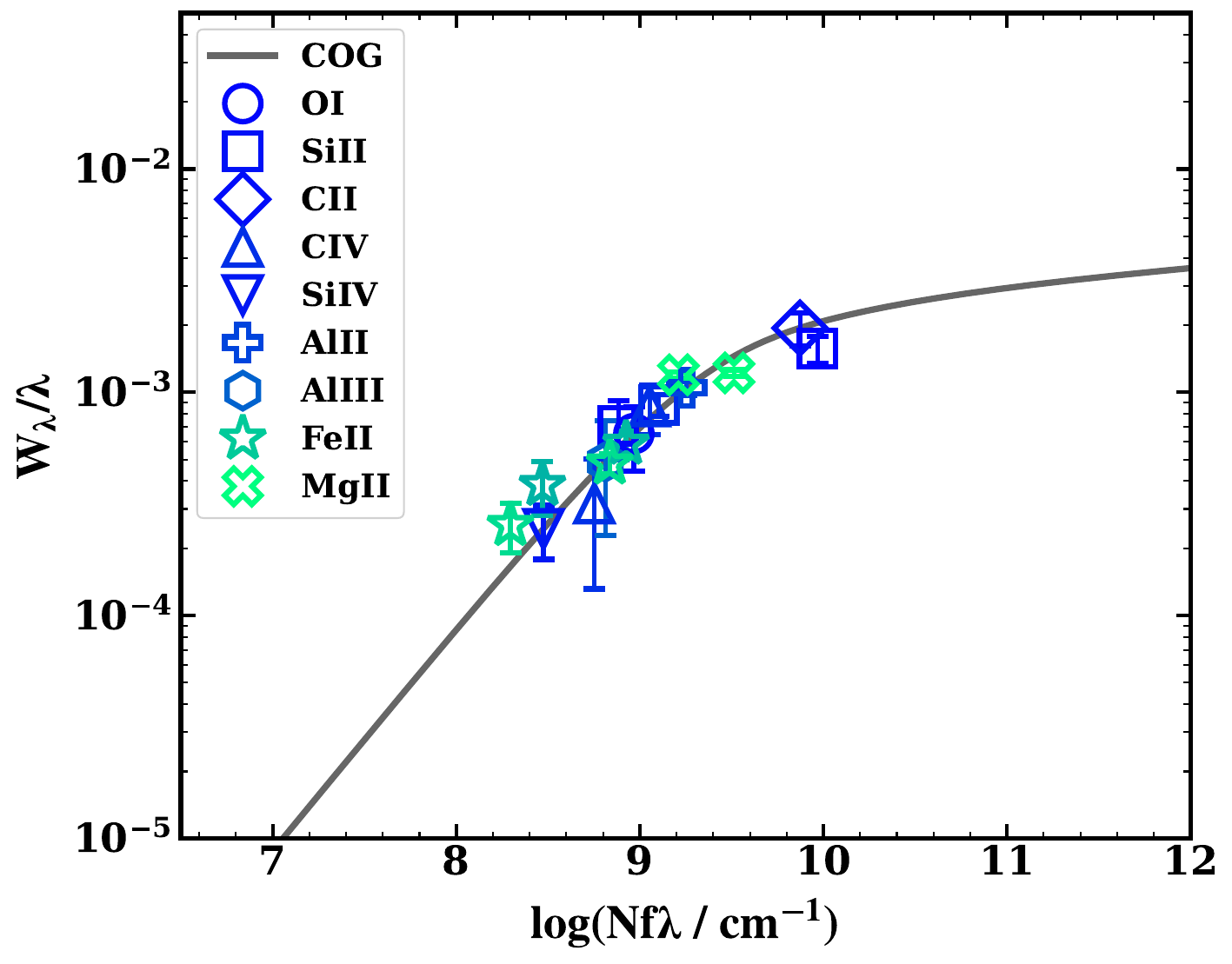}
    \caption{
    Multi-ion single-velocity-component curve-of-growth (COG) analysis for the absorption system toward GNz7q, assuming a fixed Doppler parameter $b=200\,\rm km\,s^{-1}$.
    The reduced coordinate on the horizontal axis is $\log(N \cdot f \cdot \lambda\,/\,\mathrm{cm}^{-1})$, where $\lambda$ is the rest-frame wavelength in cm.
    Data points show the rest-frame $W_\lambda/\lambda$ for each detected transition, color-coded and symbol-coded by ion species, with error bars reflecting the measurement uncertainties.
    The solid black curve is the theoretical Gaussian-profile COG computed at $b=200\,\rm km\,s^{-1}$.
    }
    \label{fig03: cog}
\end{figure}

\begin{figure*}
    \centering
    \includegraphics[width=\linewidth]{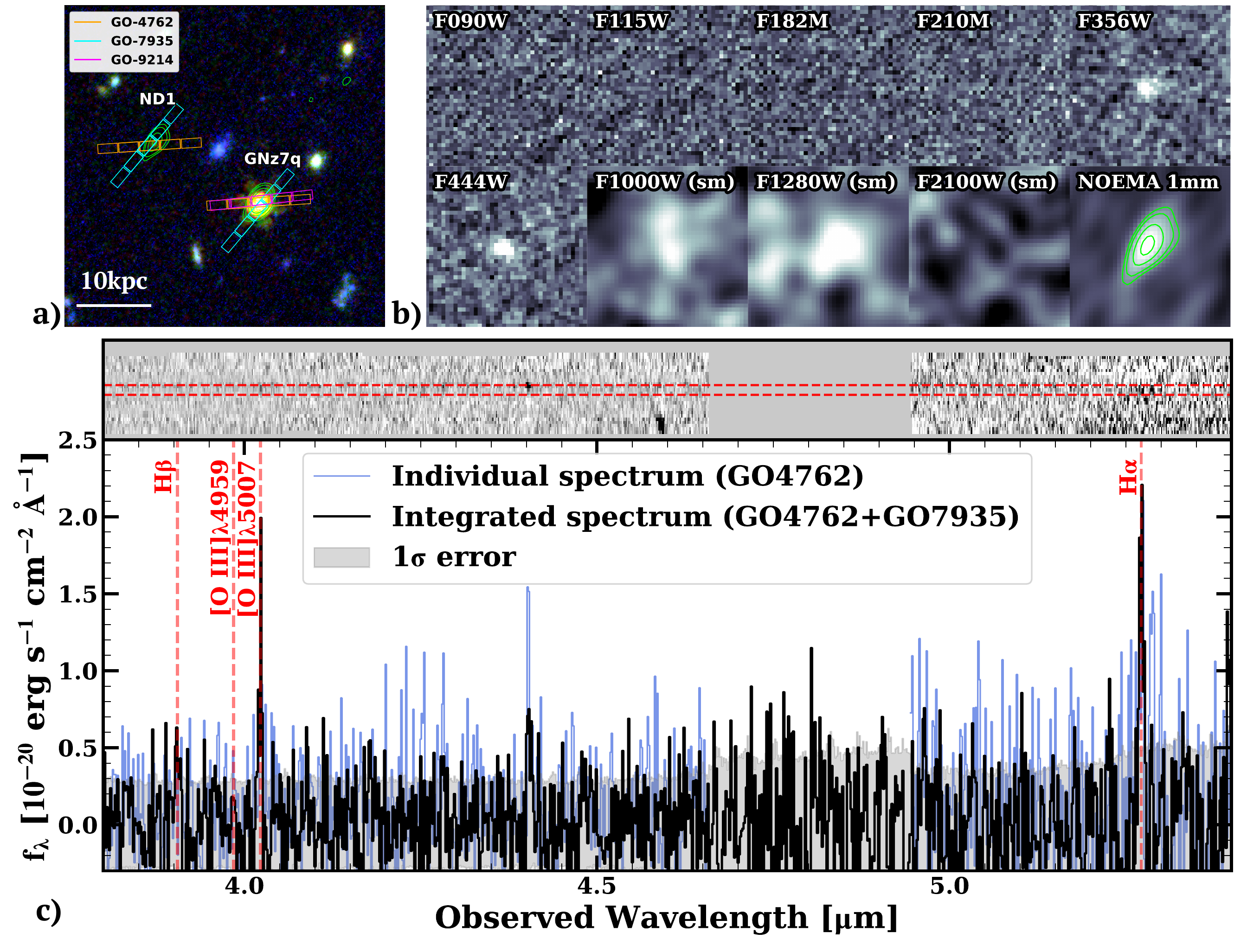}
    \caption{
    {Multi-wavelength characterization of the galaxy \targ.}
    (a) The false-color image centered on \tarq, generated from F115W, F356W, and F444W, alongside the NIRSpec MSA shutters of the three spectroscopic programs used in this study (GO~4762, GO~7935, and GO~9214; Table~\ref{tab:data}).
    The target of this study, \targ, is located toward the North--East.
    The green contours show the NOEMA 1\,mm dust continuum from the archival Band-3 observations of \citet{Fujimoto+2022}.
    (b) Multi-band cutout images of \targ, spanning near-infrared bandpasses (F090W through F2100W) and the 1\,mm NOEMA continuum map (from \citealt{Fujimoto+2022}). Each cutout is $2''\times2''$ ($\approx10\times10$\,pkpc at $z=7.03$). MIRI images are smoothed with a 2-pixel kernel for display purposes.
    (c) The JWST/NIRSpec G395M-F290LP spectrum of \targ. The top panel shows the 2D spectrum from GO\#4762, and the bottom panel shows the 1D spectrum in units of $10^{-19}\,\mathrm{erg\,s^{-1}\,cm^{-2}\,\text{\AA}^{-1}}$.
    The black line shows the integrated 1D spectrum combining two observations, while the blue line shows the 1D spectrum solely from GO\#4762.
    Key emission lines detected in the spectrum, including \textsc{[O\,iii]}$\lambda$5007 and H$\alpha$ (red labels), are marked by red dashed vertical lines, and provide the redshift anchor for the systemic velocity used in the rest of the analysis.
    }
    \label{fig04:host}
\end{figure*}

\begin{figure*}
    \centering
    \includegraphics[width=\linewidth]{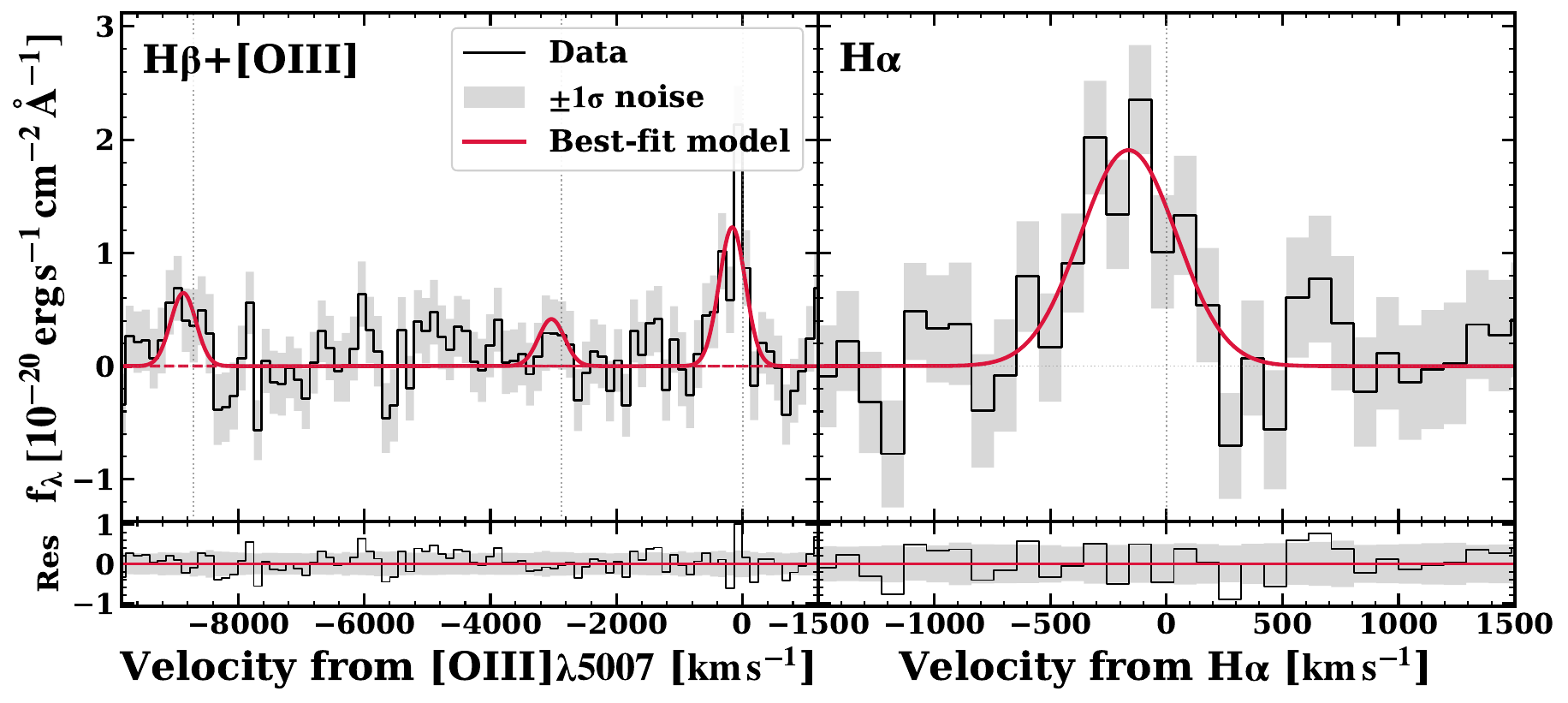}
    \caption{
    {Joint emission-line fit to the deep combined NIRSpec/G395M spectrum of ND1.}
    Left: continuum-subtracted \hb+[\ion{O}{3}]$\lambda\lambda$4959, 5007 window, with the velocity axis referenced to [\ion{O}{3}]$\lambda$5007.
    Right: continuum-subtracted spectrum and the best-fit model around \ha.
    In each panel, the observed spectrum (black histogram) and its $\pm1\sigma$ noise (grey band) are overplotted with the best-fit joint model (red solid).
    }
    \label{fig05:opt lines}
\end{figure*}

\begin{figure*}
    \centering
    \includegraphics[width=\linewidth]{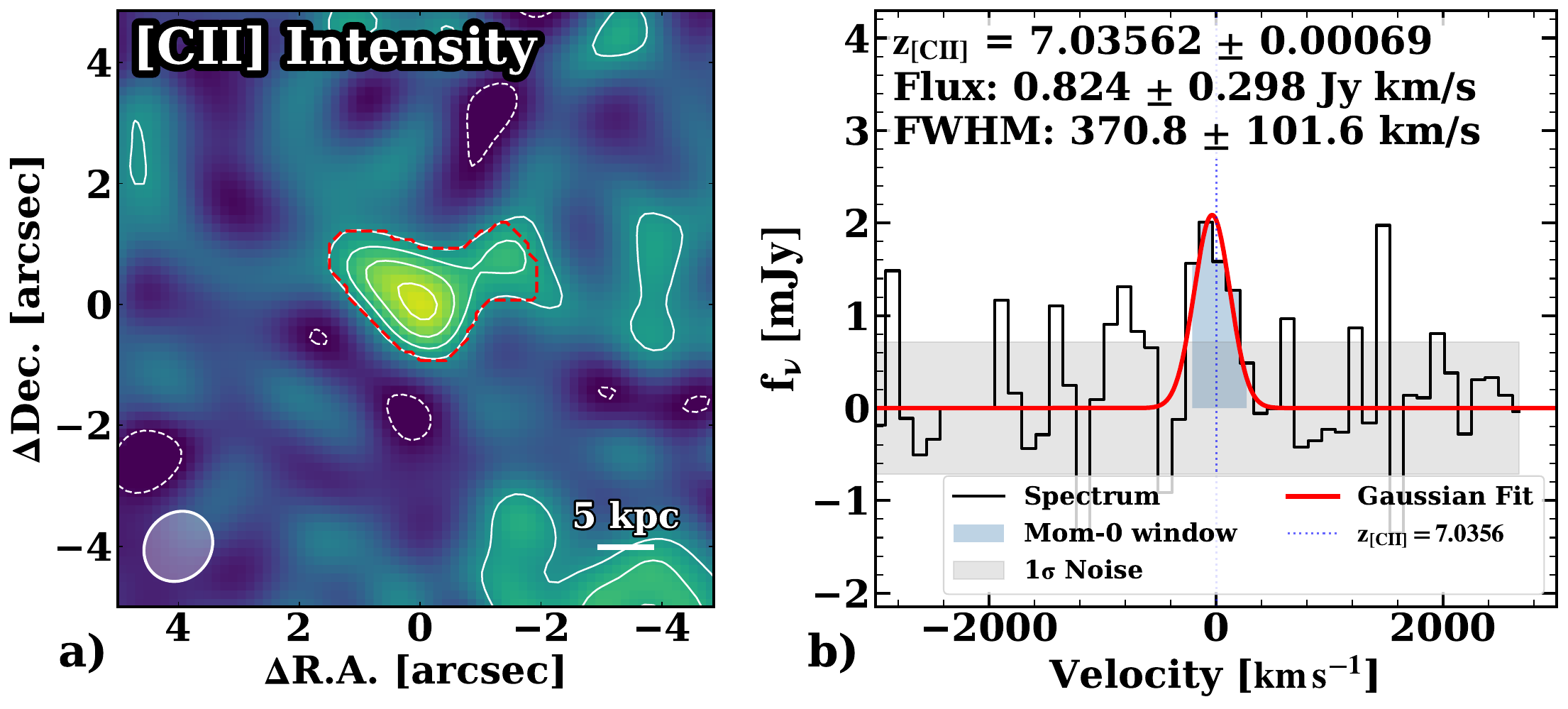}
    \caption{
    {NOEMA observations of the \ciiemi\ line in \targ.}
    (a) Velocity-integrated intensity map of \targ.
    White contours correspond to $-2, 2, 3, 4$, and $5\sigma$, where $\sigma$ represents the rms noise of the map.
    The synthesized beam and a physical scale bar are indicated in the bottom-left and bottom-right corners, respectively.
    The red dashed contour outlines the extraction footprint used to build the 1D spectrum: pixels above the $2\sigma$ moment-0 mask within a $2''$ radius around the peak.
    (b) Extracted 1D spectrum from the data cube centered on the \ciiemi\ transition. The observed data are shown in black, with the $1\sigma$ uncertainty denoted by the gray shaded region.
    The solid red curve shows the best-fit Gaussian profile, and the vertical blue line marks the fitted centroid of the \ciiemi\ line, $z_{\rm [CII]} = 7.03562$; the blue shaded region marks the channels used to construct the moment-0 map.
    }
    \label{fig05:NOEMA}
\end{figure*}

 We derive gas-phase column densities from the measured equivalent widths using a curve-of-growth (COG) analysis. We perform a multi-ion single-component COG fit to the measured equivalent widths ($W_\lambda$), i.e., we assume the absorbing gas can be represented by a single kinematic component sharing the same Doppler parameter $b$ across all detected ions. The optical depth is modeled as:
\begin{align}
    \tau_\nu=\tau_0\exp\left[-\left(\frac{u}{b}\right)^2\right], \quad b\equiv \sqrt{2}\sigma_v,
    \label{eq2: tau}
\end{align}
where $u \equiv c(\nu_0-\nu)/\nu_0$, and $\sigma_v$ is the velocity dispersion.
The line-center optical depth, $\tau_0$, is given by:
\begin{align}
    \tau_0=0.7580\left(\frac{N_l}{10^{13}\,\rm cm^{-2}}\right) \left(\frac{f_{lu}}{0.4164}\right) \left(\frac{\lambda}{1215.7\,\text{\AA}}\right) \nonumber \\ \left(\frac{10\,\rm km\,s^{-1}}{b}\right),
    \label{eq3: tau_0}
\end{align}
where the subscript $l$ denotes the lower ionic level of the transition, $N_l$ is the column density of the absorbing ion in the given transition, $f_{lu}$ is the oscillator strength for absorption from the lower level $l$ to the upper level $u$ of the transition, and $\lambda$ is the rest-frame wavelength of the transition (both $f_{lu}$ and $\lambda$ are taken from \citealt{Morton+2003_atom}).
Because the metal transitions we detect are kinematically coincident to within the measurement uncertainties, we assume ions of similar ionization potential trace a common absorbing region and share the same $b$, allowing all species to be fit simultaneously in a single multi-ion COG (Figure~\ref{fig03: cog}).
The equivalent width of each transition follows from integrating $1-e^{-\tau_\nu}$ over the line profile (Equation~\ref{eq1: EW}), so that Equations~\ref{eq1: EW}--\ref{eq3: tau_0} provide a direct mapping between $W_0$ and $N$ for a given $b$.

 For fixed atomic data, the COG has a single free physical parameter, the Doppler parameter $b$ (Equation~\ref{eq2: tau}).
Physically, $\sigma_v$ is the internal broadening of an individual absorbing cloud produced by thermal motion of the ions together with the small-scale turbulent and micro-kinematic velocity field within that cloud.
Because the NIRSpec G140M grating has an instrumental FWHM of $\approx 300\,\rm km\,s^{-1}$ ($\sigma_{\rm LSF}\approx 130\,\rm km\,s^{-1}$; \citealt{Jakobsen+2022_NIRSpec, deGraaff+2024}), which is much broader than the $\sigma_v \lesssim 100\,\rm km\,s^{-1}$ typical of individual high-$z$ CGM clouds \citep{Davies+2023_met_abs}, the intrinsic $b$ of any single kinematic component is not directly recoverable from these data.

Rather than adopt an arbitrarily narrow $b$ that would place the strong transitions on the flat or damping-wing branches of the COG, where $N$ rises steeply for a given $W_\lambda$, we deliberately fix $b = 200\,\rm km\,s^{-1}$ throughout our analysis.
This choice is not a physical claim about the internal dynamics of any single cloud.
Rather, it is an intentionally conservative operational value that is larger than any plausible internal velocity dispersion of an individual CGM cloud, and that places every detected transition on the linear or linear-to-flat portion of the COG.
In this regime, the mapping from the measured $W_\lambda$ to the inferred $N$ is monotonic, only weakly $b$-dependent, and, crucially, \emph{minimizes} $N$ for any given EW.
Any smaller and more physically motivated internal Doppler parameter (e.g., $b \sim 10$--$50\,\rm km\,s^{-1}$ for individual unresolved clouds) would push the strong transitions into the flat and damping-wing regimes and would only \emph{increase} the inferred $N$ by up to several dex (see the sensitivity analysis in Appendix~\ref{app:cog_b} and Figure~\ref{fig16:cog_b}).
Our fiducial choice therefore cannot bias the results in a direction that would artificially inflate the absorber's column densities.
The values reported in Table~\ref{tab:absorption} should instead be interpreted as conservative lower limits on the true intrinsic $N$, and the column densities of the absorber, which already lie $\approx1$--1.5\,dex above the median of $z>5$ absorbers (Section~\ref{sec4.1}), would only increase under any more realistic assumption on the internal gas kinematics.

 Two caveats accompany this single-component approximation.
First, most quasar sightlines can traverse several discrete clouds with heterogeneous column densities and kinematic widths.
The adopted $b$ is then biased by the velocity dispersion of the strongest saturated components rather than by the intrinsic thermal/turbulent broadening of the mass-bearing phase.
Second, when several narrow, saturated components are blended into a single artificially broad profile, the COG method further underestimates the intrinsic column density \citep{Prochaska+2006_GRB}.
Both effects reinforce our earlier point that the reported $N$ values are conservative lower limits.
To quantify the sensitivity of $N$ to the assumed $b$ value, we performed a systematic investigation spanning a range of Doppler parameters.
This analysis and its implications are presented in Appendix~\ref{app:cog_b}.

\subsection{Host Galaxy}
\label{host}

\begin{figure*}
    \centering
    \includegraphics[width=\linewidth]{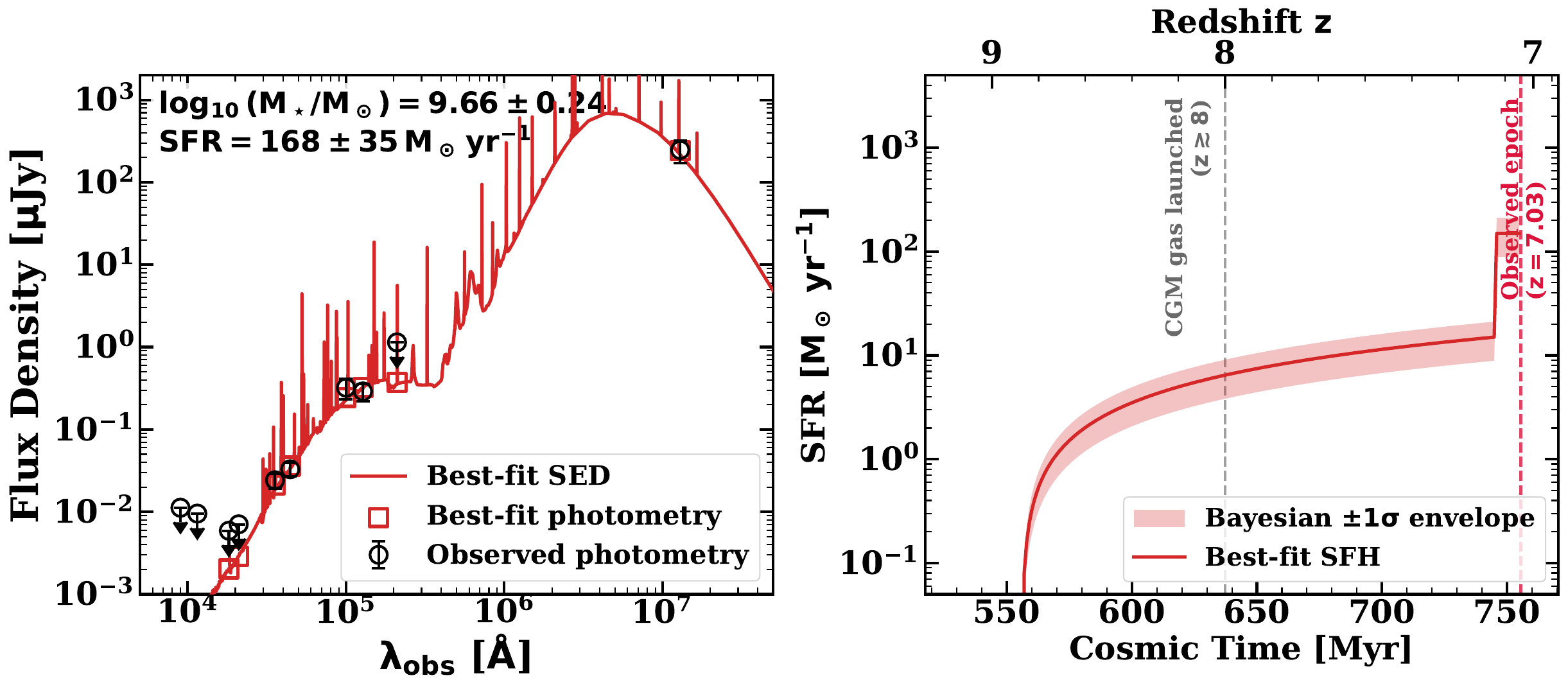}
    \caption{
    Best-fit SED and star formation history (SFH) of \targ\ from \cigale.
    Left: best-fit SED. The red solid line shows the best-fit model.
    Observed photometry is shown as black circles, with upper limits as downward-pointing triangles; open red squares mark the synthetic photometry from the best-fit model. The Bayesian stellar mass and instantaneous SFR are annotated in the top-left corner.
    Right: best-fit SFH (red solid line) with the $1\sigma$ Bayesian envelope on ${\rm SFR}_{100}$ shown as the red shaded band. Vertical dashed lines mark the inferred epoch of CGM gas ejection ($z\gtrsim 8$, gray) and the observed redshift ($z=7.03$, crimson). The $\sim 100\,\text{Myr}$ interval between them is consistent with the travel time required for the outflowing gas to reach the observed impact parameter of $\sim 16\,\text{kpc}$.
    }
    \label{fig06:sed_sfh}
\end{figure*}

\begin{figure*}
    \centering
    \includegraphics[width=\linewidth]{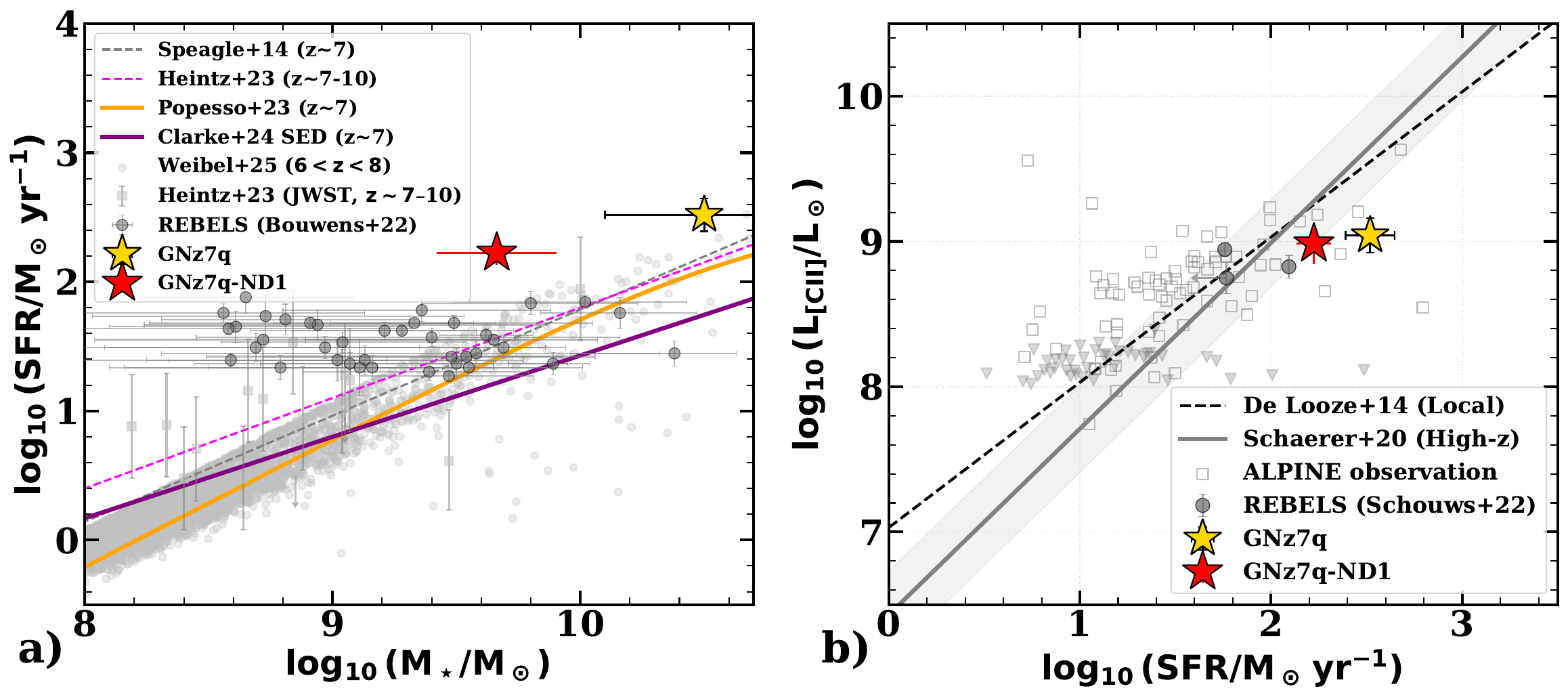}
    \caption{
    (a) Star formation rate (SFR) as a function of stellar mass ($M_*$).
    The red star represents \targ.
    Main-sequence relations from the literature are shown as colored lines, including pre-JWST results \citep{Speagle+2014_ms, Popesso+2023_ms} and new JWST results \citep{Heintz+2023_ms, Clarke+2024_ms}.
    Literature measurements of the host galaxies themselves (i.e., from their own emission) are shown as grey squares for the JWST spectroscopic sample of \citet{Heintz+2023_ms}, small light-grey dots for the JWST photometric sample at $6<z<8$ of \citet{Weibel+2024_ms}, and dark-grey circles for the REBELS sample \citep{Bouwens+2022_REBELS}.
    The host galaxy of the quasar is also shown for reference \citep[yellow star;][]{Fei+2026_GNz7q}.
    (b) \ciiemi$158\ \mu\text{m}$ luminosity ($L_{\rm [CII]}$) versus SFR.
    \targ\ is indicated by the red star.
    The black dashed and solid grey lines represent empirical relations derived from previous studies \citep{DeLooze+2014_CII, Schaerer+2020_CII}.
    Literature data are represented by open grey squares for the ALPINE sample \citep{LeFevre+2020_ALPINE, Bethermin+2020_ALPINE} and dark-grey circles for the REBELS sample \citep{Schouws+2023_REBELS}, with upper limits denoted by grey triangles.
    }
    \label{fig07:SFR}
\end{figure*}

We have successfully identified the host galaxy associated with these metal absorption lines.
This system was originally reported as a millimeter-bright source discovered via NOEMA 1-mm continuum mapping \citep[][Figure~\ref{fig04:host}]{Fujimoto+2022}.

\targ~was observed with JWST/NIRSpec in two programs (PIDs: GO\#4762 and GO\#7935; Figure~\ref{fig04:host}a).
In GO\#7935, the 2D spectrum of \targ\ is contaminated by the trace of the much brighter quasar \tarq, which overlaps with that of \targ\ on the detector in this MSA configuration.
A clean spectrum of \targ\ was therefore obtained by modeling and subtracting the quasar contribution ({\it F.~Sun, priv. comm.} ).
We then combined the spectra from the two programs to obtain a higher-S/N spectrum for \targ.
This combined spectrum yields $5\sigma$ detections of \oiiiemi\ and \ha\ and only a marginal ($3\sigma$) detection of \hb\ (Figure~\ref{fig04:host}c).
We do not detect [\ion{O}{3}]$\lambda$4959, consistent with the theoretical line ratio of $f_{\rm [OIII]\lambda5007}/f_{\rm [OIII]\lambda4959}=2.98$ \citep{Storey+2000_OIII}.

We jointly fit the \hb, [\ion{O}{3}]$\lambda\lambda$4959,5007 and \ha\ emission lines in the deep combined NIRSpec/G395M spectrum of ND1 with a common velocity offset and intrinsic velocity width, and the [\ion{O}{3}] doublet ratio fixed at the value of 1:2.98 \citep{Storey+2000_OIII}.
The best-fit model is shown in Figure~\ref{fig05:opt lines}.
The fit yields a redshift of $z= 7.0312 \pm 0.0008$ and best-fit line fluxes of $F_{\rm H\alpha} = (1.74 \pm 0.34) \times 10^{-18}$, $F_{\rm H\beta} = (4.4 \pm 1.5) \times 10^{-19}$, and $F_{\rm [OIII]\lambda 5007} = (8.6 \pm 1.7) \times 10^{-19}\,\rm erg\,s^{-1}\,cm^{-2}$.
   The Balmer decrement is $F({\rm H\alpha})/F({\rm H\beta}) = 4.0 \pm 1.4$, which we convert into a dust attenuation assuming the \cite{Calzetti+2000} law and an intrinsic ratio of 2.86.
We convert the \ha\ luminosity into an SFR with the calibration of \citet{Kennicutt+2012_ARAA}, which traces star formation over the past $\approx10$\,Myr. The dust-uncorrected value is ${\rm SFR}_{\rm H\alpha}=5.4\pm1.1\,M_\odot\,{\rm yr^{-1}}$. Because the Balmer decrement is measured at low S/N, we propagate the uncertainties of both line fluxes with a Monte Carlo analysis, imposing the physical prior $F({\rm H\alpha})/F({\rm H\beta})\geq2.86$ (i.e., $A_V\geq0$). This yields $A_V=1.1^{+1.5}_{-1.1}$\,mag and a dust-corrected ${\rm SFR}_{\rm H\alpha}=13^{+31}_{-8}\,M_\odot\,{\rm yr^{-1}}$ (median and 16th--84th percentiles).

Complementary NOEMA observations also detected the \ciiemi$\lambda$158$\rm \mu m$ fine-structure line, with a peak signal-to-noise ratio of $5\sigma$ in the velocity-integrated moment-0 map (Figure~\ref{fig05:NOEMA}).
Fitting the \ciiemi\ line alone yields a redshift of $z=7.03562 \pm 0.00069$, with a flux of $f_{\rm [CII]}=0.824\pm0.298\,\rm Jy\,km\,s^{-1}$ and a line width of $\rm FWHM_{[CII]}=370.8\pm101.6\,km\,s^{-1}$, yielding a \ciiemi\ luminosity of $\log (L_{\rm [CII]}/L_\odot)=9.01\pm0.16$.

We derive the systemic redshift by combining the optical and far-infrared lines as follows.
We fit a single Gaussian profile (plus a constant) independently to the velocity profiles of \oiiiemi$\lambda5007$ and \ha\ in the GO~4762 G395M spectrum and of \ciiemi\ in the NOEMA spectrum, and take the inverse-variance-weighted mean of the three velocity centroids.
This yields $z_{\rm sys}=7.0321\pm0.0007$, which we adopt as the fiducial redshift of \targ\ in the following analysis.
The \ciiemi-only redshift is offset from the optical one by $\approx170\,{\rm km\,s^{-1}}$ ($\approx1.4$ NOEMA channels), which we include as a systematic uncertainty in the kinematic comparison of Section~\ref{sec5.1}.

 Photometrically, the source is undetected in all JWST/NIRCam short-wavelength (SW) filters (F090W to F210M, i.e., rest-frame wavelengths up to $\approx2600$\,\AA\ at $z=7.03$; $3\sigma$ upper limits of 6--11\,nJy), whereas it is detected in the long-wavelength (LW) bands F356W and F444W at ${\rm S/N}\approx5$ (Figures~\ref{fig04:host}b and \ref{fig06:sed_sfh}).
Furthermore, it is marginally detected in the MIRI F1000W and F1280W filters with ${\rm S/N}\approx3.6$ and 4.2, respectively.
However, it is not detected in the F2100W band (Figure~\ref{fig06:sed_sfh}, left).
This non-detection is attributable to the shallower depth of that observation (Table~\ref{tab:data}): its $3\sigma$ upper limit of $1.1\,\mu$Jy lies above the flux density predicted by the best-fit SED ($\approx0.4\,\mu$Jy).
We then perform spectral energy distribution (SED) fitting for \targ\ with the \cigale\ software package \citep{Boquien+2019_cigale, Yang+2022_cigale}, using the ten broad-band photometric measurements available for the source.
We hold the redshift fixed at $z = 7.0321$ (the value derived above from the joint [OIII], H$\alpha$, and [CII] fit, with a statistical uncertainty of $\sigma_z = 7\times 10^{-4}$), and all available UV-to-millimeter photometric measurements are incorporated into the fit.
For the stellar emission, we adopt the \cite{BC03} single stellar population synthesis models (implemented in \cigale\ as the \texttt{bc03} sub-module of the stellar-population library) alongside a \cite{Chabrier2003_IMF} initial mass function (IMF).
We parameterize the star formation history (SFH) using a delayed-$\tau$ model supplemented by a recent exponential starburst component, with the mass fraction of this late starburst allowed to vary between 0 and 0.5 in steps of 0.1.
We choose this two-component SFH because it is flexible enough to reproduce both a smoothly rising early SFH and the strongly obscured, burst-like activity indicated by the very red UV--MIR continuum, while remaining compact enough (five free parameters) to be well constrained by the ten photometric measurements. We verified that simpler prescriptions (single delayed-$\tau$, single constant SFH, or a non-parametric ``sfh\_nonparam'' history) yield stellar masses and SFRs consistent with our fiducial values within their $1\sigma$ uncertainties.
Following the methodology of \cite{Fei+2026_GNz7q}, we constrain the maximum age of the stellar population to be less than 95\% of the age of the Universe at the source redshift ($t_{\rm age}\leq 0.95\,t_{\rm H}$).
Nebular emission is incorporated via \texttt{nebular}.
To account for the potentially extreme interstellar medium (ISM) conditions typical of high-redshift galaxies, the ionization parameter ($\log U$) is allowed to vary between $-3$ and $-1$, with gas-phase metallicities ranging from 10\% to 100\% of the solar metallicity.
Dust attenuation is modeled using a modified \cite{Calzetti+2000} attenuation law (\texttt{dustatt\_modified\_starburst}).

The resulting best-fit SED yields a stellar mass of $\log (M_*/M_\odot)=9.66\pm0.24$ and an instantaneous star formation rate (SFR) of $168\pm35\,M_\odot\,\rm yr^{-1}$ (the 100\,Myr-averaged SFR is ${\rm SFR_{100Myr}}=26\pm10\,\rm M_\odot\,yr^{-1}$).
The best-fit SED and the corresponding SFH are shown in the left and right panels of Figure~\ref{fig06:sed_sfh}, respectively.
This measured SFR places the host galaxy approximately $2\sigma$ above the star-forming main sequence at this epoch \citep[panel a of Figure~\ref{fig07:SFR};][]{Speagle+2014_ms, Heintz+2023_ms, Popesso+2023_ms, Clarke+2024_ms}.

Comparing this measured $L_{\rm [CII]}$ to the star formation rate (SFR) derived from our SED fitting shows that \targ\ is well aligned with established empirical $L_{\rm [CII]}$--SFR scaling relations (right panel of Figure~\ref{fig07:SFR}; \citealt{DeLooze+2014_CII, Schaerer+2020_CII}). The location of \targ\ on the SFR--$M_\star$ main sequence relative to comparable high-$z$ galaxy samples is also shown in the left panel of Figure~\ref{fig07:SFR}.
The SED-based SFRs exceed the dust-corrected \ha\ estimate by a factor of $\approx13$ for the instantaneous SFR and $\approx2$ for ${\rm SFR}_{100}$. Part of this difference reflects the different timescales probed: \ha\ traces star formation over the past $\approx10$\,Myr, whereas the dust continuum and the SED fit are sensitive to star formation over $\approx100$\,Myr \citep{Kennicutt+2012_ARAA}.
More importantly, the Balmer decrement only corrects the attenuation of the regions from which \ha\ escapes, so the difference indicates that a large fraction of the star formation in \targ\ is optically thick in the rest-frame optical and visible only in the dust continuum.
Collectively, these physical properties indicate that the host galaxy, \targ, is consistent with a star-forming galaxy at $z \sim 7$, lying $\approx2\sigma$ above the main sequence.

\begin{deluxetable}{lc}
\tablecaption{Physical properties of \targ.\label{tab:host}}
\tablewidth{0pt}
\tablehead{
  \colhead{Quantity} & \colhead{Value} %
}
\startdata
$\rho$ & $16$\,kpc \\ %
$z_{\rm sys}$ & $7.0321 \pm 0.0007$ \\ %
$z_{\rm opt}$ & $7.03117 \pm 0.00073$ \\ %
$z_{\rm [CII]}$ & $7.03562 \pm 0.00069$ \\ %
$\log(M_*/M_\odot)$ & $9.66 \pm 0.24$ \\ %
SFR & $168 \pm 35\,M_\odot\,{\rm yr}^{-1}$ \\ %
${\rm SFR}_{\rm 100Myr}$ & $26 \pm 10\,M_\odot\,{\rm yr}^{-1}$ \\ %
$A_V$ & $1.1^{+1.5}_{-1.1}$\,mag \\ %
$\rm SFR_{\rm H\alpha}$ & $13^{+31}_{-8}\,M_\odot\,{\rm yr}^{-1}$ \\ %
$f_{\rm [CII]}$ & $0.824 \pm 0.298$\,Jy\,km\,s$^{-1}$ \\ %
$\rm FWHM_{[CII]}$ & $370.8 \pm 101.6\,{\rm km\,s^{-1}}$ \\ %
$\log(L_{\rm [CII]}/L_\odot)$ & $9.01 \pm 0.16$ \\ %
$r_{\rm vir}$ & $\approx 22^{+10}_{-7}\,{\rm kpc}$ \\ %
$v_{\rm esc}$ & $\approx 220\,{\rm km\,s^{-1}}$ \\ %
\enddata
\tablecomments{Uncertainties are $1\sigma$. Column densities and outflow parameters of the associated absorber are listed in Table~\ref{tab:absorption}.}
\end{deluxetable}

\section{Results}
\label{sec4}
  \subsection{Metal-rich halo gas with high column densities}
 \label{sec4.1}

Quantitatively, our curve-of-growth (COG) analysis yields column densities on the order of $\log (N/\rm cm^{-2}) \approx 14$--$15$ across all detected species (Table~\ref{tab:absorption}).
We stress at the outset that these values are conservative lower limits on the true intrinsic column densities in this system, as detailed in Section~\ref{sec3.2}.
However, even under this conservative choice, the recovered column densities are among the highest reported at these epochs.
Our \ion{C}{2} column density, $\log(N_{\rm C\,II}/{\rm cm^{-2}})\geq15.6$, exceeds the largest \ion{C}{2} column density ($15.1$) among the $z>6$ absorbers of the E-XQR-30 survey \citep{Davies+2023_met_abs}.
Whereas typical high-$z$ metal-absorber studies span $10^{12}$--$10^{14}\,\rm cm^{-2}$, our measurements lie above the median of the $z>5$ absorbers of \citet{Davies+2023_met_abs} by 1.55\,dex for \ion{C}{2}, 1.4\,dex for \ion{Si}{2}, 1.25\,dex for \ion{Mg}{2}, 1.1\,dex for \ion{Fe}{2}, 1.2\,dex for \ion{C}{4} (a lower limit), and 0.7\,dex for \ion{O}{1}, which places \ion{Fe}{2} and \ion{Si}{2} at the high end of the known distribution.
The qualitative interpretation of an unusually dense, chemically enriched halo gas is therefore not sensitive to the specific choice of $b$.
Any physically motivated lower $b$ would only result in even larger column densities (Appendix~\ref{app:cog_b}).

\begin{figure}
    \centering
    \includegraphics[width=\linewidth]{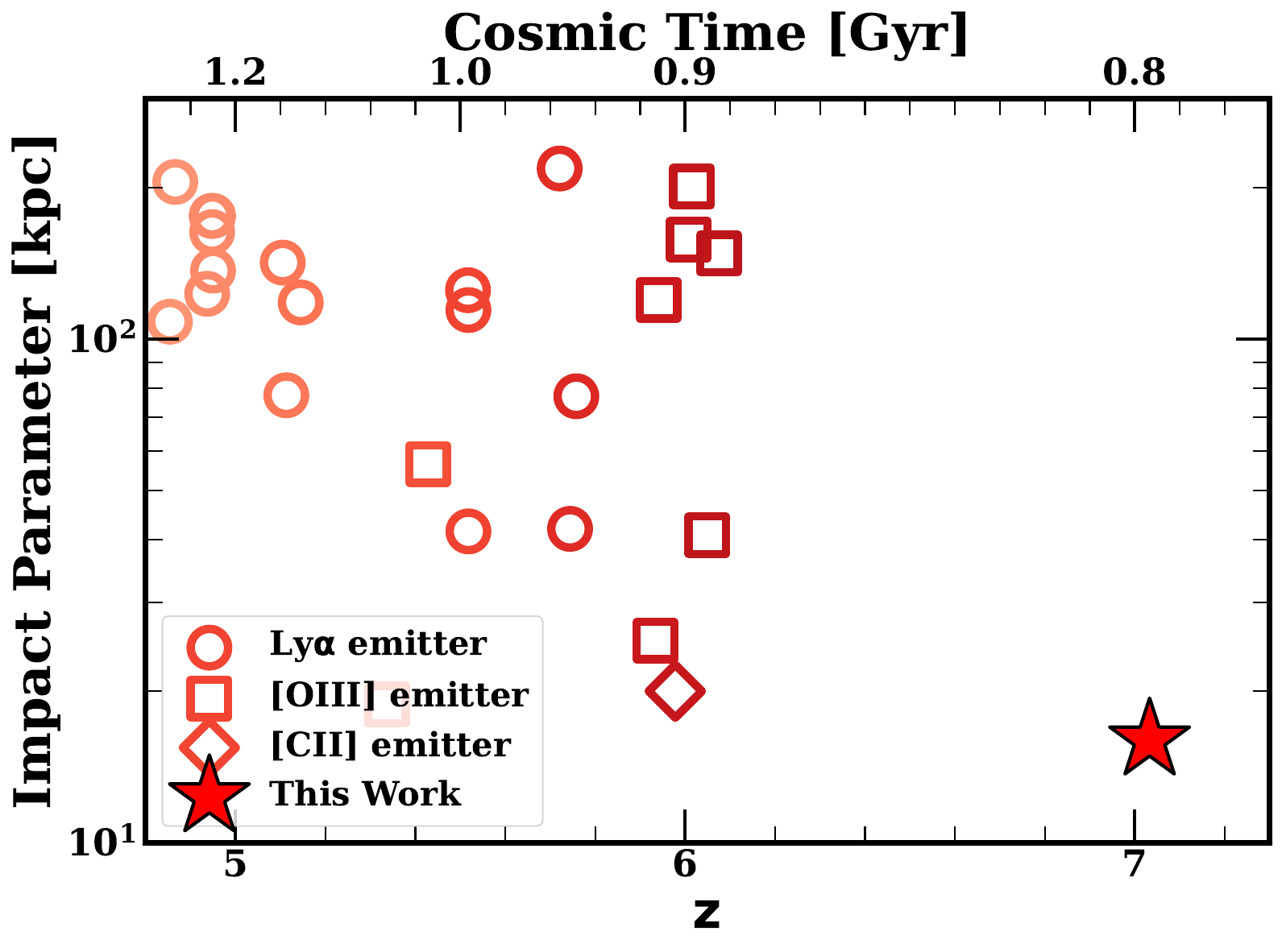}
    \caption{
    Impact parameter versus redshift for high-$z$ ($z\gtrsim 5$) galaxy--absorber pairs.
    \targ\ is shown as the red star, and the literature sources are shown as red open symbols, with different symbols indicating different tracers for the identification of galaxies \citep[circles for Ly$\alpha$, squares for \oiiiemi, and diamonds for \ciiemi; ][]{Wu+2021_nat, Wu+2023_MgII_abs, Diaz+2014, Diaz+2015, Diaz+2021, Cai+2017, Bordoloi+2024_EIGER, Sebastian+2026_CGM}.
    }
    \label{fig09:impact}
\end{figure}

\begin{figure*}
    \centering
    \includegraphics[width=\linewidth]{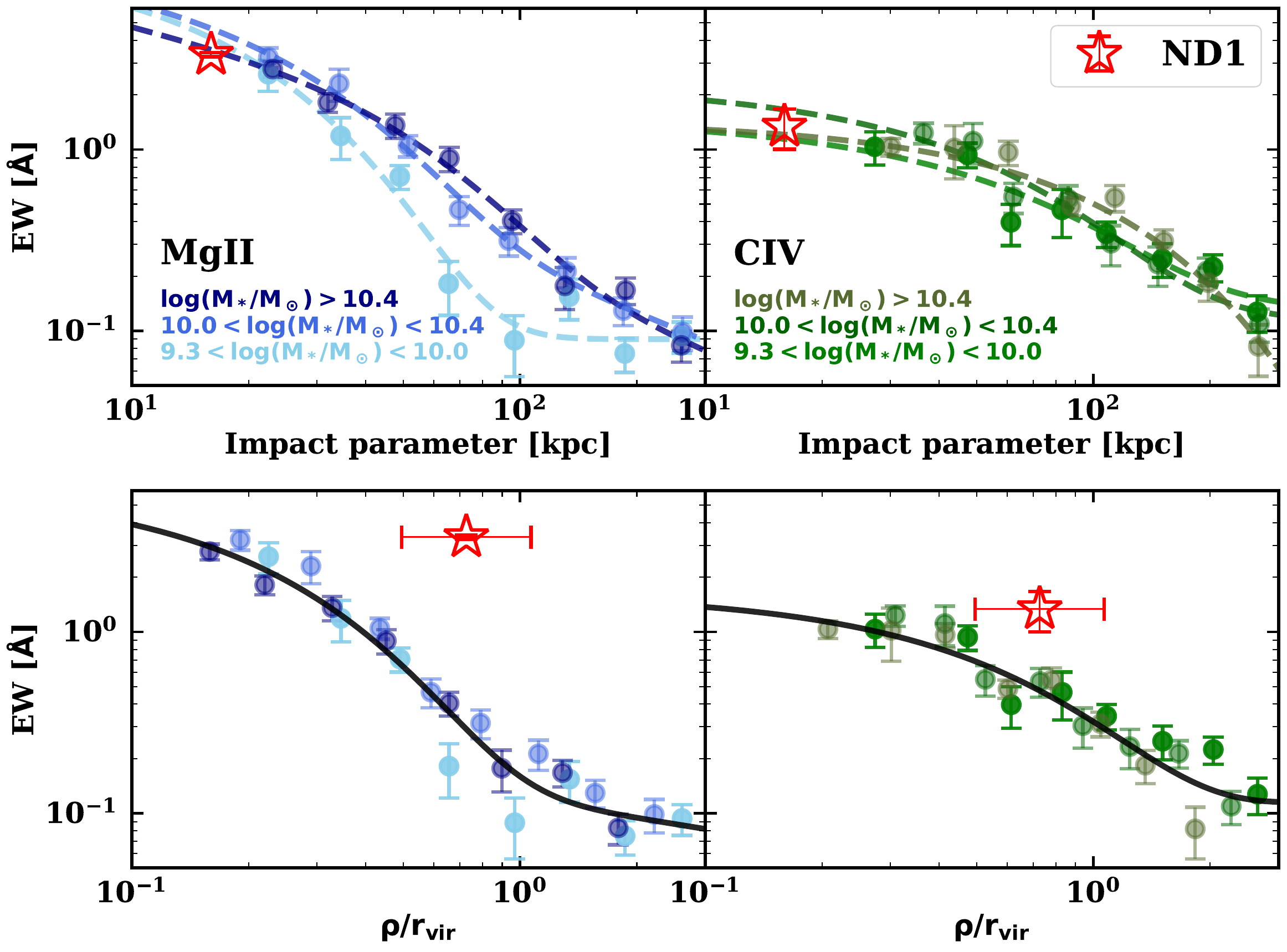}
    \caption{
    Top panel: observed rest-frame equivalent widths of absorption lines as a function of impact parameter as extracted from this study (red star) and the DESI survey \citep[solid blue and green circles; ][]{Lan+2025_DESI}.
    The left and right panels present the \ion{Mg}{2}\ and \ion{C}{4}\ absorption lines, respectively.
    Bottom panel: EWs vs. virial radius ($r_{\rm vir}$)-normalized impact parameter ($\rho/r_{\rm vir}$). Symbols are the same as in the top panel.
    The horizontal error bar in the bottom panels shows a conservative $\pm0.5$\,dex uncertainty in the inferred halo mass ($\log M_h/M_\odot\approx 11.3$ from the stellar-to-halo mass relation of \citealt{Behroozi+2019_SAM}), corresponding to $r_{\rm vir}\approx15$--32\,kpc. The low-redshift data and relations are measured for galaxies with known stellar masses (CGM absorption toward background quasars), as is ND1.
    }
    \label{fig12:ew}
\end{figure*}

In addition to analyzing the absorption column density, we compare the impact parameter ($\rho$) of our system against established galaxy-absorber pairs in the literature.
As shown in Figure~\ref{fig09:impact}, our target presents the smallest impact parameter observed to date ($\rho=16\,\rm kpc$), $\approx35\%$ smaller than the typical literature measurements at these redshifts ($\rho\sim 25\,\rm kpc$).

 Specifically, an impact parameter of $16\,\rm kpc$ falls well within the anticipated virial radius of the host galaxy's dark matter halo.
Using the empirical galaxy-halo mass relation from \cite{Behroozi+2019_SAM}, we obtain a rough dark matter halo mass of $M_h\approx10^{11.3}\,M_\odot$, with a corresponding virial radius of $r_{\rm vir}\approx22$\,kpc. Allowing for a conservative $\pm0.5$\,dex uncertainty in $M_h$, which encompasses the scatter of the stellar-to-halo mass relation and the uncertainty in $M_*$, gives $r_{\rm vir}\approx15$--32\,kpc.
 Consequently, the background quasar sightline intersects the innermost regions of the CGM.

We further compare the $W_0$ of the \ion{Mg}{2}\ and \ion{C}{4}\ transitions to the established empirical $W_0$--$\rho$ (top panel of Figure~\ref{fig12:ew}) and $W_0$--$\rho/r_{\rm vir}$ (bottom panel) relations derived from comprehensive statistical analyses of the low-to-intermediate-redshift CGM \citep[$z\sim 1-2$; e.g.,][]{Nielsen+2013_MgII, Bordoloi+2014_CIV, Lan+2025_DESI}.
On the $W_0$--$\rho$ plane, both transitions in \targ\ lie within $\lesssim1\sigma$ of the low-redshift $\log(M_*/M_\odot)\sim 10$ track at the small-$\rho$ end, and the measured $W_0$ magnitudes are consistent with the extrapolated relations to $\rho\approx 16$\,kpc.
On the normalized $W_0$--$\rho/r_{\rm vir}$ plane, the position of \targ\ moves up and to the right relative to the low-redshift trend.
   The horizontal error bar in Figure~\ref{fig12:ew} shows the conservative $\pm0.5$\,dex range in $M_h$, which corresponds to $\rho/r_{\rm vir}\approx0.50$--1.07. Over this full range, \targ\ lies a factor of $\approx5$--20 (\ion{Mg}{2}) and $\approx2$--4.5 (\ion{C}{4}) above the best-fit low-redshift relations, so the excess on the normalized plane is not an artifact of the uncertain $r_{\rm vir}$. Because $r_{\rm vir}$ shrinks rapidly with redshift at fixed halo mass, this excess suggests that the high-redshift CGM at a given fraction of the virial radius contains more metal-enriched gas than at $z\sim1$--2, although a single object cannot establish such a conclusion.
A single target precludes strong statistical conclusions about the global evolution of the CGM absorption strength.
Nevertheless, the agreement in physical impact parameter suggests that the galactic feedback mechanisms responsible for building the enriched inner CGM operate in a broadly similar way across cosmic time.

\subsection{Metallicity \& Ionization State}
\label{sec4.4:rat}

\begin{figure*}
    \centering
    \includegraphics[width=\linewidth]{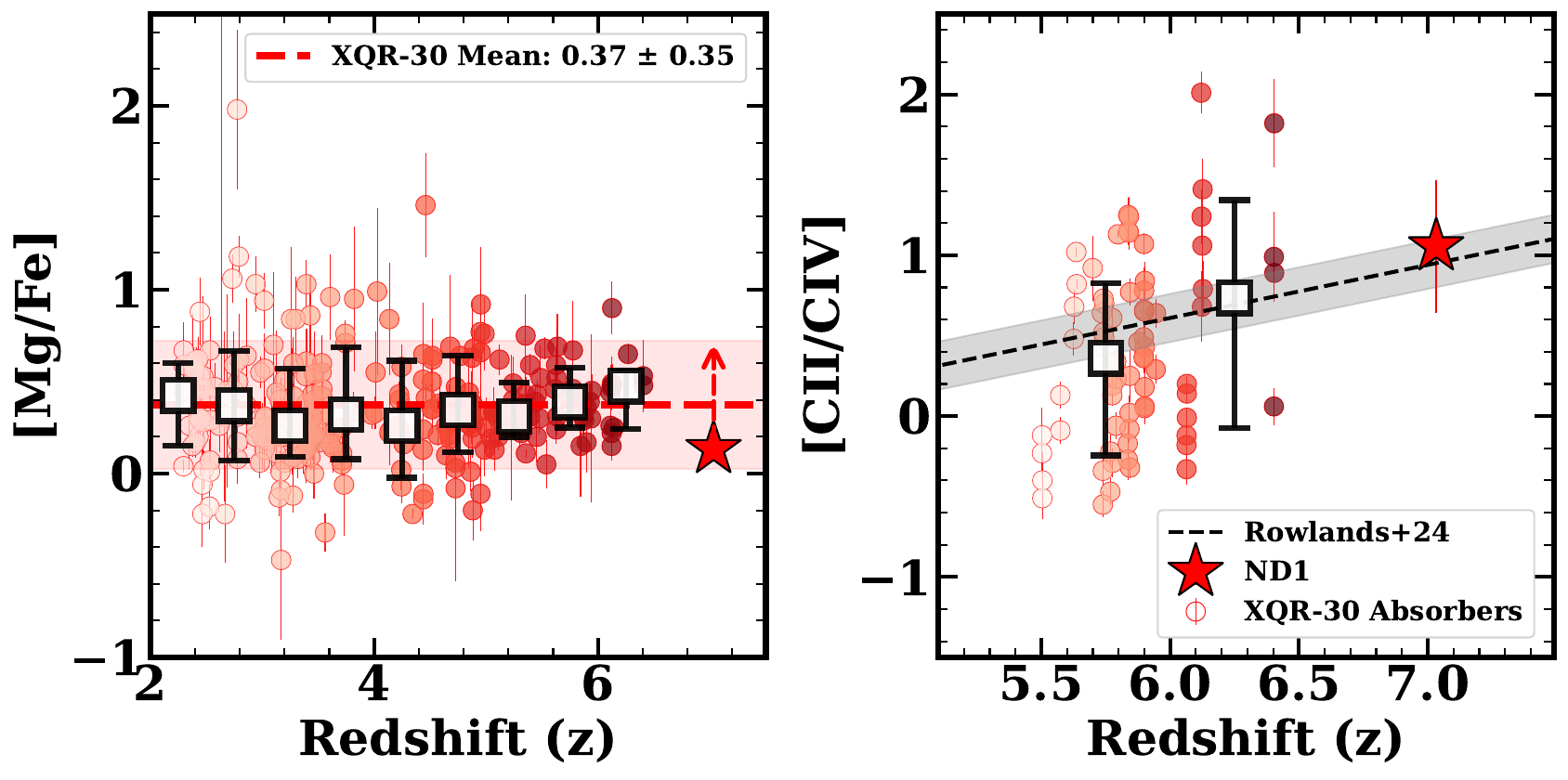}
    \caption{
    {Redshift distribution of the abundance ratios in the CGM of \targ.} \targ\ is shown as the red star in both panels.
    Left: abundance ratio of \ion{Mg}{2} to \ion{Fe}{2}.
    The measurement for \targ\ is compared to an archival sample from the literature \citep{Davies+2023_met_abs, Rowlands+2026_EXQR}, with the mean of the literature distribution indicated by the red dashed line; both \targ\ and the literature sample adopt the solar ratio of \citet{Asplund+2009}, and the dashed arrow indicates that dust depletion of Fe biases the observed [Mg/Fe] of \targ\ upward.
    Right: abundance ratio of \ion{C}{2}\ to \ion{C}{4}, compared to literature measurements shown as red circles.
    The empirical redshift-dependent trend of the \ion{C}{2}/\ion{C}{4}\ ratio is overlaid as a black dashed line, while the grey shaded region represents the 0.3 dex dispersion of this relation \citep{Rowlands+2026_EXQR}.
    }
    \label{fig11:z_evo}
\end{figure*}

Utilizing the column densities derived from our curve-of-growth analysis, we then estimate the abundance ratio between \ion{Mg}{2} and \ion{Fe}{2}\ ([Mg/Fe]$\equiv \log (N_{\rm Mg}/N_{\rm Fe})-\log (N_{\rm Mg}/N_{\rm Fe})_\odot$).
We estimate this specific line ratio because it serves as a standard diagnostic for the nucleosynthetic history.
Magnesium is a classic $\alpha$-element produced rapidly by massive stars via core-collapse supernovae, whereas iron is predominantly synthesized by Type Ia supernovae on longer, multi-hundred-million-year timescales.

In Figure~\ref{fig11:z_evo}, we examine whether the relative [Mg/Fe] abundance ratio changes with redshift, comparing the \targ\ absorber against the comprehensive XQR-30 legacy sample of intervening systems \citep{Davies+2023_met_abs}.
   The redshift interval sampled here ( $z\sim2$--$7$) spans $\approx2.5$\,Gyr of cosmic time (from $\approx0.75$ to $\approx3.2$\,Gyr after the Big Bang). Because each data point corresponds to a different absorber rather than to the same gas observed at different epochs, any trend should be interpreted as a change of the absorber population with redshift rather than as chemical evolution in the strict sense.
Over the redshift range $z \sim 2$--$6$, the literature sample demonstrates a constant mean of $[\rm Mg/Fe] = 0.37 \pm 0.35$.
 We convert the ionic column densities into an abundance ratio as
\begin{equation}
[{\rm Mg/Fe}] = \log\left(\frac{N_{\rm Mg\,II}}{N_{\rm Fe\,II}}\right) - \log\left(\frac{\rm Mg}{\rm Fe}\right)_\odot,
\label{eq:mgfe}
\end{equation}
where we assume that \ion{Mg}{2} and \ion{Fe}{2} are the dominant ionization states of Mg and Fe in the low-ionization gas (both have ionization potentials of $\approx15$--16\,eV for the next ionization stage, so no differential ionization correction is applied), and adopt the solar value $\log({\rm Mg/Fe})_\odot=7.60-7.50=0.10$ \citep{Asplund+2009}.
 Using the column densities from Table~\ref{tab:absorption}, $\log N_{\rm Mg\,II}=14.277^{+0.019}_{-0.021}$ and $\log N_{\rm Fe\,II}=14.043^{+0.038}_{-0.045}$, we measure $[\rm Mg/Fe] = +0.13 \pm 0.05$ for \targ.
This value is consistent with the literature mean at $z\sim2$--6 within its $0.35$\,dex dispersion.

However, we need to be cautious that the observed [Mg/Fe] ratio may be heavily modulated by dust depletion, given that the host galaxy of the CGM is a heavily dust-obscured galaxy.
In this case, a substantial fraction of its circumgalactic iron is likely locked into solid dust grains (\ion{Mg}{2} is moderately depleted, and \ion{Fe}{2} is heavily depleted), artificially inflating the observed gas-phase [Mg/Fe] ratio \citep{Jenkins+2009}.

We also examine the column density ratio of \ion{C}{2}/\ion{C}{4}, which can be used to diagnose the ionization state of the absorbing gas and the shape of the local radiation field.
In Figure~\ref{fig11:z_evo}, we present the \ion{C}{2}/\ion{C}{4} abundance ratio for \targ, alongside literature samples from the XQR-30 survey and the empirical relation \citep[e.g., ][]{Davies+2023_met_abs, DOdorico+2023_XQR30, Rowlands+2026_EXQR}.
Previous studies demonstrate a clear evolutionary trend, where the \ion{C}{2}/\ion{C}{4} ratio systematically increases at higher redshifts \citep{Rowlands+2026_EXQR}. In principle, such an increase can arise either from a decline of \ion{C}{4} or from an increase of \ion{C}{2}. The observed absorber statistics indicate that both can contribute.
The incidence of \ion{C}{4} absorbers declines toward $z\sim6$ \citep[e.g.,][]{Becker+2009_CIV, DOdorico+2023_XQR30}, whereas that of low-ionization absorbers such as \ion{O}{1} and \ion{C}{2} increases \citep{Becker+2019_OI, Cooper+2019_CGM}, pointing to a change in the ionization state of the metal-enriched gas rather than in its total metal content.
Because the \ion{C}{4} column density is a lower limit (Table~\ref{tab:absorption}), the measurement for \targ, $\log(N_{\rm C\,II}/N_{\rm C\,IV})\leq1.06$, is an upper limit. This upper limit lies within the 0.3\,dex standard deviation of this established empirical relationship at $z=7.033$ (right panel of Figure~\ref{fig11:z_evo}).
This consistency suggests that despite its high column densities ($\gtrsim1$\,dex above typical) and heavily dust-obscured host, the overall ionization balance of the CGM surrounding this system follows the broader cosmic evolution of early galactic halos.

   The strong \ion{C}{4} absorption of \targ \ therefore does not require an unusually hard radiation field, and can instead be a natural consequence of the large total metal column density along this sightline.
While a massive, starburst-driven outflow might be expected to heavily ionize the surrounding halo, thereby suppressing the \ion{C}{2}/\ion{C}{4}\ ratio below the high-redshift average, our data show no significant deviation from the broader early-Universe population.

\section{Discussion}
\label{sec5}

The star formation history (SFH) of \targ, while subject to large uncertainties due to the limited photometric constraints available at $z \sim 7$, encodes important information about the timing of early chemical enrichment.
The right panel of Figure~\ref{fig06:sed_sfh} presents the best-fit SFH, which indicates sustained star formation extending to $z > 8$, as required to explain the observed CGM enrichment.
We first discuss the constraints on the CGM ejection epoch from outflow kinematics, emphasizing the inherent projection uncertainties (Section~\ref{sec5.1}), and then examine the chemical abundance ratios as a fossil record of early Population~III enrichment (Section~\ref{sec5.3}).

\subsection{Gas Outflow and the CGM/IGM enrichment}
\label{sec5.1}

Before interpreting the kinematics, we briefly outline the possible physical origins of the observed \tarq--\targ\ absorber and the arguments that guide our preferred interpretation.
Given the projected separation ($\rho\approx 16$\,kpc), the close alignment of the absorber redshift with the systemic redshift of \targ\ (\(|\Delta z|\lesssim 3\times10^{-3}\) between the metal-line and \oiiiemi/H\(\alpha\)/\ciiemi\ centroids), and the multi-species, high-column-density character of the absorption, the plausible origin scenarios are:

(i) a metal-rich galactic outflow launched from \targ\ that intercepts the \tarq\ sightline;

(ii) cold accretion or a recycling galactic fountain feeding gas back onto \targ;

(iii) tidal or ram-pressure-stripped material from a companion galaxy; or

(iv) chance alignment with an unrelated intervening system.

The negligible line-of-sight velocity offset ($\lesssim 100\,\rm km\,s^{-1}$) between the absorber and \targ\ makes an unrelated chance alignment or a strongly infalling stream disfavored, and the very high column densities are difficult to sustain in pristine accreting gas.
In addition, we do not detect any close companions around \targ.
Therefore, scenarios (iii) and (iv) are very unlikely.
Scenarios (i) and (ii) are similar in that both imply that the metal-enriched gas was ejected by star formation in \targ.
In the remainder of this section, we focus on the implications for star-formation feedback, assuming scenario (i).

\begin{figure*}
    \centering
    \includegraphics[width=0.9\linewidth]{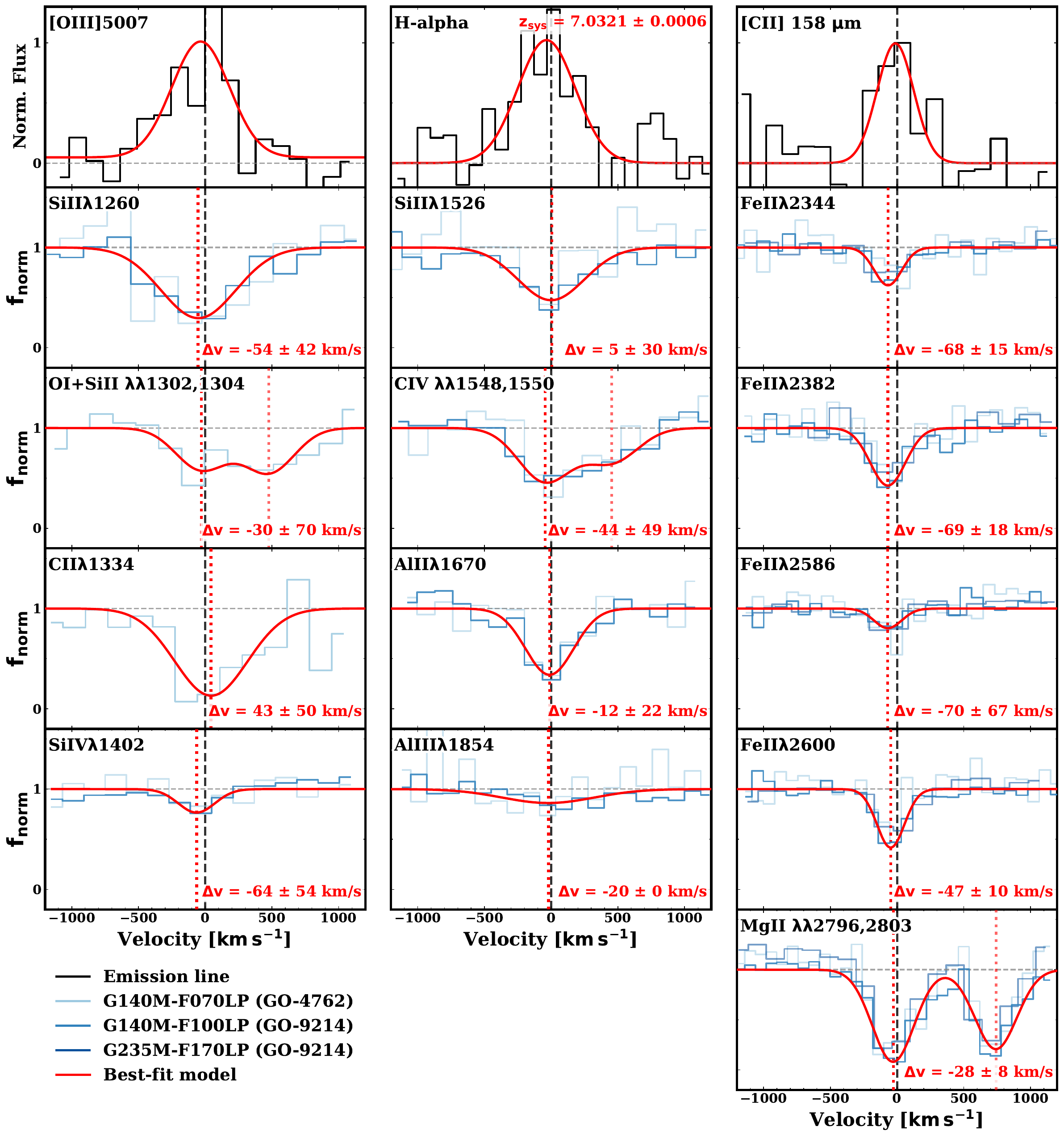}
    \caption{
    Comparison between the emission lines of \targ\ and the absorption lines measured from the background quasar spectrum.
    The $x$-axis shows the line-of-sight velocity relative to the systemic redshift of the host galaxy, as derived from the \oiiiemi\ and $\rm H\alpha$ emission lines.
    The observed spectra are presented as black histograms, with the best-fit Gaussian profiles overlaid as solid red curves.
    Red dotted vertical lines indicate the centroids of the individual absorption components, whereas the black vertical dashed line marks the systemic velocity.
    }
    \label{fig10:kin}
\end{figure*}

We also compare the velocity center of the absorbing gas and the host galaxy (Figure~\ref{fig10:kin}).
By modeling the absorption lines with Gaussian profiles and measuring the velocity centroids of the absorbing gas, we derive velocity shifts of $\sim0$--$100\,\rm km\,s^{-1}$ (Figure~\ref{fig10:kin}).
This offset is consistent with typical outflow velocities observed in the local Universe and at high redshifts \citep{Heckman+2000_outflow, Veilleux+2005_review, Zhu+2026_feedback}.

Both $\rho$ and $\Delta v$ are projected quantities, where $\rho$ is the on-sky separation and thus a lower limit of the true 3D distance ($r\geq\rho=16$\,kpc), while $\Delta v$ is the line-of-sight velocity component and thus a lower bound on the intrinsic outflow speed. The intrinsic ratio $v/r$ is therefore not preserved under projection in general, and the values reported below should be read as first-order estimates (see Appendix~\ref{app:outflow_geom} for a discussion of how these projection biases enter the outflow-rate estimate).
To incorporate the underlying kinematic dispersion of the outflowing gas, and thereby recover a more physically meaningful characterization of the wind speed, we follow the prescription of \citet{Fiore+2017_AGN_of} and define the outflow velocity as
\begin{equation}
v_{\rm out} = |\Delta v| + 2\sigma_{\rm int},
\end{equation}
where $\Delta v$ is the fitted centroid offset from the systemic redshift and $\sigma_{\rm int}$ is the intrinsic Gaussian width of the absorption profile, obtained by deconvolving the NIRSpec MSA line spread function ($\rm FWHM_{LSF}\approx250\,\rm km\,s^{-1}$, $\sigma_{\rm inst}\approx106\,\rm km\,s^{-1}$) from the fitted width in quadrature \citep{deGraaff+2024}.
Averaging this quantity across the low-ionization transitions in our sample yields a characteristic outflow velocity of $v_{\rm out}\approx300\,\rm km\,s^{-1}$, with individual transitions spanning $\sim 250$--$570\,\rm km\,s^{-1}$.
The maximum blueshifted extent is delivered by the strong resonance lines (\ion{Si}{2}$\lambda1260$, \ion{C}{2}$\lambda1334$), which, owing to their large oscillator strengths, remain detectable in low-column-density, high-velocity gas and therefore probe the fastest-moving gas in the outflow.

 The outflow velocities differ substantially among ions, from $|v_{\rm out}|\approx50$--70\,km\,s$^{-1}$ for the weaker \ion{Fe}{2} transitions to $\approx550$--680\,km\,s$^{-1}$ for \ion{Si}{2}$\lambda1260$, \ion{C}{2}, and \ion{Al}{3} (Table~\ref{tab:absorption}). This spread may well be physical. At the NIRSpec resolution, each measured profile is a column-density-weighted blend of all clouds along the sightline, and a galactic wind is expected to consist of multiple clouds with different velocities, densities, and ionization states. Transitions with different ionization potentials and oscillator strengths weight these clouds differently.
 Strong low-ionization lines saturate in the densest, slowest clouds and remain sensitive to tenuous, fast-moving clouds, whereas weak transitions (e.g., \ion{Fe}{2}$\lambda\lambda2344,2586$) are dominated by the densest component. The ion-dependent velocities are therefore consistent with a multi-cloud, multiphase outflow, although measurement systematics (line blending and the $\approx170$\,km\,s$^{-1}$ uncertainty in the systemic redshift; Section~\ref{host}) also contribute. Resolving the individual clouds will require spectroscopy at $R\gtrsim10\,000$.

 Assuming a typical stellar-to-halo mass ratio and standard concentration \citep{Behroozi+2019_SAM, Dutton+2014_DMC}, we derive a circular velocity of $\sim 150\,\rm km\,s^{-1}$ and a corresponding escape velocity at the projected 16\,kpc impact parameter of $v_{\rm esc}\approx 220\,\rm km\,s^{-1}$.
Given that the outflow velocities ($\approx300$\,km\,s$^{-1}$) are comparable to, or even larger than, the escape velocity $v_{\rm esc}$, and noting that the velocity measurement carries systematic channel-to-channel noise of order $\sim$100\,km\,s$^{-1}$, the comparison should be treated as suggestive rather than conclusive.
With this caveat, the results are at minimum consistent with gas that can permeate and extend throughout the host halo, naturally explaining the discovery of metal-enriched extended gas at these epochs \citep{Fujimoto+2019_halo}, and potentially consistent with material that can escape to enrich the surrounding IGM.

Adopting this outflow velocity as the characteristic launch speed of the enriched gas, we can place first-order constraints on the timescale of early CGM enrichment.
Traversing the projected $16\,\rm kpc$ separation at $v_{\rm out}\approx300\rm\,km\,s^{-1}$ requires a minimum travel time of $\Delta t \gtrsim 50\,\rm Myr$.
Given the host systemic cosmic age of $t_{\rm sys}\simeq 753\,\rm Myr$ (corresponding to $z_{\rm sys}=7.0321$), this places the launching episode at $z \gtrsim 7.5$, an epoch of active star formation in the host inferred from the SED reconstruction (Figure~\ref{fig06:sed_sfh}, right panel) and consistent with the compact, dust-obscured burst phase of ND1.
This estimate should be regarded as a lower limit.
Correcting the projected separation for an inclination factor of $\sim 2$ ($r \sim 30\,\rm kpc$) doubles the required travel time to $\Delta t \sim 100\,\rm Myr$, pushing the launching epoch back to $z \gtrsim 8$.
If, moreover, the absorbing gas traces material embedded in a recycling galactic fountain, in which the outflowing wind is decelerated by the halo potential and subsequently re-accreted, the launching epoch should be even earlier to accommodate the additional deceleration and turnaround timescales.

\subsection{Mass loading}
\label{sec5.2}

Assuming that the \ion{Si}{2}-absorbing gas is representative of the outflowing wind and treating the CGM absorber as a partially covering thin shell at the projected impact parameter $\rho$, we estimate the mass outflow rate following the method of \citet{Rupke+2005, Chisholm+2017} (see Appendix~\ref{app:outflow_geom} for the full derivation and a sketch of the assumed geometry),
\begin{equation}
\dot{M}_{\rm out} = 4\mu m_{\rm H} C_f N_{\rm H} \rho v_{\rm out},
\end{equation}
where $\mu = 1.4$ accounts for the mean mass weight in the presence of helium, $m_{\rm H}$ is the mass of a hydrogen atom, $C_f \simeq 0.5$ is an empirical covering fraction typical of high-$z$ intervening absorbers (at higher redshift, $C_f$ can be larger; $C_f\simeq 0.6$ from \citealt{Nakane+2026_che}, while here we adopt a conservative value).
$N_{\rm H}$ is the total hydrogen column density along the sightline, and $v_{\rm out}$ is the outflow velocity derived in the previous section.
Since Si{\sc ii} is the dominant ionization state of silicon in cool, moderately dense CGM gas, no ionization correction is required, and we adopt $N_{\rm Si}\simeq N_{\rm SiII}$.
The hydrogen column density is then recovered as
\begin{equation}
\log N_{\rm H} = \log N_{\rm SiII} - [\rm Si/H] - \log\left(N_{\rm Si}/N_{\rm H}\right)_\odot,
\end{equation}
with solar abundances from \citet{Asplund+2009} and a fiducial CGM metallicity of $[\rm Si/H] = -1.5$, consistent with the $[\rm Si/O]\sim+0.8$, $[\rm O/H]\sim-2$ enrichment pattern characteristic of $z>6$ QSO absorbers \citep{Christensen+2023_qso_abs, Sodini+2024_che}.
Adopting our measured \ion{Si}{2} column density $\log N_{\rm SiII} = 14.80^{+0.07}_{-0.08}$ (see Table~\ref{tab:absorption}) yields $N_{\rm H} \approx 6\times 10^{20}\,\rm cm^{-2}$, which, in combination with $\rho = 16\,\rm kpc$ and $v_{\rm out} = 300\,\rm km\,s^{-1}$, gives
\begin{equation}
\dot{M}_{\rm out} \approx 64\,M_\odot\,{\rm yr}^{-1}.
\end{equation}
Accounting for the dominant systematic uncertainties, namely the assumed metallicity ($-2 \leq [\rm Si/H] \leq -1$), the covering fraction ($C_f = 0.3$--$1.0$), and the intrinsic outflow velocity, gives a plausible range of $\dot{M}_{\rm out}\sim23$--$260\,M_\odot\,{\rm yr}^{-1}$.

Combining this with the 100-Myr-averaged SFR from the panchromatic \cigale\ fit, $\mathrm{SFR}_{100} = 26 \pm 10\,M_\odot\,{\rm yr}^{-1}$, we infer a mass-loading factor of
\begin{equation}
\eta \equiv \frac{\dot{M}_{\rm out}}{\mathrm{SFR}} \approx 3_{-2}^{+7}.
\end{equation}
Such a high mass loading characterizes strong, energy-driven galactic winds expected in compact, dust-obscured starbursts \citep[e.g.,][]{Muratov+2015_FIRE, Nelson+2019_TNG50}, and is larger than the values inferred from Mg{\sc ii} absorption studies of massive $z\sim1$--$2$ star-forming galaxies \citep[$\eta\sim0.5$--$3$;][]{Rupke+2005, Weiner+2009_DEEP2}.
However, this result is broadly consistent with the increasingly efficient mass loading measured toward higher-redshift, lower-mass systems in recent JWST outflow studies \citep[e.g., ][]{Asada+2026_glimpse}, further supporting the physical picture in which the CGM of \targ\ has been chemically enriched by a sustained, energetic feedback episode originating from a dust-obscured, burst-dominated host.

\begin{figure*}
    \centering
    \includegraphics[width=\linewidth]{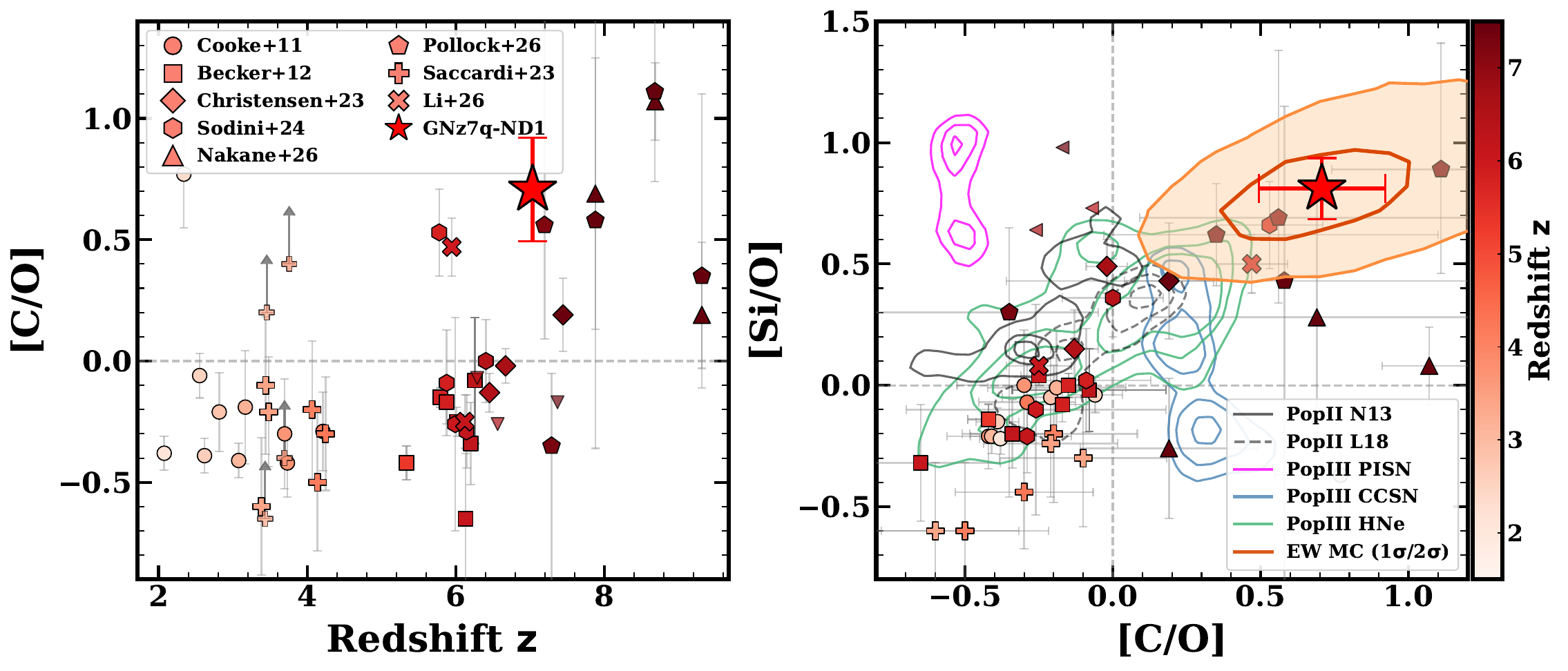}
    \caption{
    Chemical abundance revealed from absorption lines at high-$z$, including quasar absorbers (CGM/IGM of foreground galaxies) from ground-based telescopes \citep[circles, squares, and hexagons;][]{Cooke+2011_chemical, Becker+2012_chemical, Sodini+2024_che}, among them absorbers at $z\approx3$--6 with candidate Pop~III signatures \citep[plus signs and crosses;][]{Saccardi+2023_che, Li+2026_che}, and from JWST \citep[diamonds;][]{Christensen+2023_qso_abs}, down-the-barrel ISM absorption in the spectra of UV-bright galaxies from JWST \citep[triangles and pentagons;][]{Nakane+2026_che, Pollock+2026_che}, and the CGM absorber of this study (star). Downward (left panel) and leftward (right panel) triangles mark [C/O] upper limits; upward arrows mark [C/O] lower limits.
    Left: The carbon-to-oxygen abundance ratio, [C/O], as a function of redshift.
    Our target is shown as the red star, while comparative samples from the literature are presented as colored data points.
    Right: [Si/O] vs. [C/O].
    The nested orange contours around \targ\ mark the $1\sigma$ and $2\sigma$ joint-confidence regions on the abundance ratios, obtained from a Monte Carlo propagation of the equivalent-width measurement uncertainties through the multi-ion curve-of-growth fit ($N_{\rm MC}=3\times 10^{4}$; see Section~\ref{sec5.3}).
    Overlaid colored tracks show theoretical nucleosynthetic yields (not measurements) from Population II (black) and three Population III scenarios: pair-instability supernovae (magenta), core-collapse supernovae (blue), and hypernovae (green).
    }
    \label{fig13:popiii}
\end{figure*}

\subsection{Chemical record of early Population~III star formation?}
\label{sec5.3}

Beyond the kinematic constraints discussed above, the chemical abundance ratios of the circumgalactic gas encode a possible fossil record of the nucleosynthetic processes that shaped the earliest stellar generations in this system \citep[e.g.,][]{Wang_IGM2012,Jaacks2018}.
To probe intrinsic enrichment independently of dust depletion, we evaluate the carbon-to-oxygen abundance ratio, ${\rm [C/O]} \equiv \log(N_{\rm C}/N_{\rm O}) - \log(N_{\rm C}/N_{\rm O})_\odot$.
Because carbon and oxygen share similar condensation temperatures and dust-depletion efficiencies, their ratio is largely insensitive to dust-induced systematic bias --- in contrast to iron-peak abundance diagnostics, which are substantially affected by the depletion onto dust grains.

The left panel of Figure~\ref{fig13:popiii} shows [C/O] as a function of redshift for \targ\ alongside comparison samples from the literature. We note that these comparison data trace different gas: the quasar absorbers \citep[DLAs and intervening metal systems;][]{Cooke+2011_chemical, Becker+2012_chemical, Saccardi+2023_che, Christensen+2023_qso_abs, Sodini+2024_che, Li+2026_che} probe gas in the CGM/IGM of foreground galaxies along random sightlines, as does \targ, whereas the JWST measurements of \citet{Nakane+2026_che} and \citet{Pollock+2026_che} are ``down-the-barrel'' absorption lines in the rest-UV spectra of UV-bright galaxies, which probe the ISM of the galaxies themselves.
The comparison sample spans $z\approx2$--9, i.e., $\approx2.7$\,Gyr of cosmic time; because each point corresponds to a different system, any apparent trend is best described as a change of the population with redshift rather than as chemical evolution.
The [C/O] ratio of \targ\ ($+0.76^{+0.38}_{-0.31}$) is $\approx0.85$\,dex above the median ($[{\rm C/O}]\approx-0.09$) and above the largest value ($+0.66$) of the $z\sim5$--6.5 quasar absorbers of \citet{Sodini+2024_che}, making \targ\ the most carbon-enhanced quasar absorber known at $z > 5$.
Rather than following the bulk of the absorber population, whose abundance ratios are broadly reproduced by standard Population~II (Pop~II) stellar evolution models, this enhancement aligns \targ\ with the abundance patterns observed in extremely metal-poor (EMP) stars in the local Universe \citep{Cayrel+2004_EMP, Beers+2005_EMP} and in the increasing number of high-redshift systems proposed to carry chemical signatures of Population~III (Pop~III) enrichment \citep[e.g.,][]{Deugenio_Carbon2024, Nakane+2026_che, Pollock+2026_che}.

To further constrain the nucleosynthetic origin of the CGM gas, we present \targ\ in the [Si/O]--[C/O] diagnostic plane and compare it with both observational samples and theoretical nucleosynthetic yield models (Figure~\ref{fig13:popiii}, right panel; directly adopted from \citealt{Christensen+2023_qso_abs}).
The overlaid tracks represent the predicted metal yields from three Pop~III nucleosynthetic channels, including pair-instability supernovae (PISNe with progenitor masses $140$--$260\,M_\odot$ shown as magenta tracks; \citealt{Heger+2002_PISN}), which are characterized by high [Si/O] and low [C/O] owing to the absence of substantial carbon-rich ejecta, core-collapse supernovae (CCSNe with progenitor masses of $10$--$100\,M_\odot$ shown as blue tracks; \citealt{Heger+2010_CCSN}), and hypernovae (HNe with progenitor masses of $10$--$100\,M_\odot$ shown as green tracks; \citealt{Kobayashi+2006_HN}). Hypernovae are core-collapse explosions of massive stars with kinetic energies of $\gtrsim10^{52}$\,erg, i.e., $\gtrsim10$ times those of normal CCSNe, possibly powered by rapidly rotating progenitors; their higher explosion energies produce more complete and incomplete Si-burning products (e.g., Si, Zn, Co) and shift the yields toward heavier elements compared with normal CCSNe \citep{Kobayashi+2006_HN, Nomoto+2013_yields}.
 Those models yield distinct abundance signatures spanning a broad range of [C/O] and [Si/O].
For reference, the expected chemical abundance patterns from standard Pop~II stellar populations are also shown as black tracks in Figure~\ref{fig13:popiii} \citep{Nomoto+2013_yields}.

Our target occupies a locus of enhanced $[\mathrm{C/O}] = +0.76^{+0.38}_{-0.31}$ and moderate $[\mathrm{Si/O}] = +0.81^{+0.21}_{-0.18}$, where the quoted uncertainties are propagated from the individual equivalent-width measurement errors (Table~\ref{tab:absorption}) through the multi-ion COG fit via a Monte Carlo test with $N_{\rm MC}=3\times10^{4}$ realizations; the resulting $1\sigma$ and $2\sigma$ joint-confidence regions are shown as the nested orange contours in Figure~\ref{fig13:popiii} (right panel). This locus is distinct from the metal absorbers identified in surveys prior to JWST \citep{Cooke+2011_chemical, Becker+2012_chemical}, while its chemical signature is consistent with recent JWST measurements at higher redshifts \citep[$z\gtrsim 7$;][]{Nakane+2026_che, Pollock+2026_che}.
Within this region of the diagnostic plane, the observed abundance pattern is broadly consistent, at the $2\sigma$ level, with Pop~III CCSNe and HNe models, while being largely different from pure Pop~II yields.
Although moderate [Si/O] values can, in principle, be reproduced by both standard Pop~II enrichment and energetic Pop~III supernovae from low-mass progenitors ($10$--$15\,M_\odot$), the large [C/O] provides the key discriminating constraint.
Such high carbon enrichment relative to oxygen is not readily achieved by canonical Pop~II stellar evolution models alone, pointing instead to a contribution from Pop~III nucleosynthetic channels.
We caution, however, that the $2\sigma$ MC contour reaches the high-$[\mathrm{C/O}]$ tail of the Pop~II yield tracks (Figure~\ref{fig13:popiii}, right panel), so a purely Pop~II origin cannot be formally excluded from the present measurement alone; a definitive separation between the Pop~III and Pop~II scenarios will require abundance-ratio determinations of higher precision than is achievable with the current medium-resolution NIRSpec data.

 A high [C/O] can arise from either an excess of carbon or a deficit of oxygen, so we also examine carbon relative to the other measured elements. Using the column densities in Table~\ref{tab:absorption} and the solar abundances of \citet{Asplund+2009}, we find $[{\rm C/Si}]\approx-0.07$, $[{\rm C/Mg}]\approx+0.53$, and $[{\rm C/Fe}]\approx+0.67$, together with $[{\rm O/Mg}]\approx-0.20$ and $[{\rm Si/O}]\approx+0.81$. Carbon is therefore enhanced by $\approx0.5$--0.7\,dex relative to O, Mg, and Fe, but not relative to Si, while O is mildly deficient relative to Mg and strongly deficient relative to Si. The high [C/O] thus reflects both a genuine carbon enhancement relative to the $\alpha$-elements O and Mg and a relatively low oxygen abundance. We caution that Fe and, to a lesser extent, Mg and Si are depleted onto dust grains, which raises the ratios relative to Fe and Mg \citep{Jenkins+2009}, and that \ion{O}{1} is measured from a single transition blended with \ion{Si}{2}$\lambda1304$, so $N_{\rm O\,I}$ carries the largest systematic uncertainty. The robust conclusion is that carbon is at least as abundant as Si and more abundant than O and Mg relative to the solar pattern, which is difficult to reproduce with Pop~II yields.

In the [Si/O]--[C/O] plane (Figure~\ref{fig13:popiii}, right), the literature measurements closest to \targ\ are the JWST down-the-barrel absorption measurements of UV-bright galaxies at $z\approx7$--9: CEERS-1019 at $z=8.68$ ($[{\rm C/O}]=1.11$, $[{\rm Si/O}]=0.89$) and EGS-z7p2 at $z=7.19$ ($[{\rm C/O}]=0.56$, $[{\rm Si/O}]=0.69$) from \citet{Pollock+2026_che}, and EGSY8p7 at $z=8.68$ ($[{\rm C/O}]=1.07$) from \citet{Nakane+2026_che}. These systems share with \targ\ a similar epoch ($\lesssim0.8$\,Gyr after the Big Bang) and a super-solar [C/O] together with an elevated [Si/O], whereas pre-JWST quasar absorbers at $z\lesssim6.5$ cluster around solar [C/O]. The key difference is the gas probed: those measurements trace the ISM of the galaxies themselves, whereas \targ\ traces CGM gas at $\rho=16$\,kpc from a dust-obscured host. The similarity suggests that the carbon-enhanced pattern seen in the ISM of early galaxies is carried into the CGM by outflows without strong dilution by pristine or differently enriched gas, and that such a pattern may be common among galaxies at $z\gtrsim7$.

Carbon-enhanced absorbers with candidate Pop~III signatures have also been reported at lower redshifts. At $z\approx3$--4, \citet{Saccardi+2023_che} identified very metal-poor ($[{\rm Fe/H}]<-2$), optically thick absorbers, three of which are carbon-enhanced and show relative abundances similar to those of very metal-poor stars in the Local Group. Where both C and O are detected, their absorbers have sub-solar $[{\rm C/O}]\approx-0.6$ to $-0.1$, while systems with only an upper limit on O reach $[{\rm C/O}]>+0.4$ (Figure~\ref{fig13:popiii}). At $z=5.945$, \citet{Li+2026_che} found that the absorber toward SDSS~J0100+2802, with $[{\rm C/O}]=+0.47\pm0.12$ and $[{\rm Si/O}]=+0.50\pm0.12$, is consistent with enrichment dominated by Pop~III supernovae and resides in a metal-poor galaxy overdensity, whereas the absorber at $z=6.122$ toward VDES~J0224$-$4711 ($[{\rm C/O}]=-0.25\pm0.11$) is consistent with Pop~II yields. Among quasar absorbers, the J0100+2802 system lies closest to \targ\ in the [Si/O]--[C/O] plane, although its [C/O] is lower by $\approx0.3$\,dex and both ratios may decrease by up to $\approx0.3$\,dex after ionization corrections \citep{Li+2026_che}. These systems indicate that gas carrying Pop~III-like enrichment patterns persists in absorbers down to $z\approx3$; \targ\ extends this population to $z\approx7$ and has the highest [C/O] among quasar absorbers at $z>5$.

 We emphasize that these findings do not imply that the host galaxy is presently dominated by Pop~III stellar populations.
Rather, the CGM gas preserves a chemical record of an earlier epoch \citep[e.g.,][]{Jeon_IGM2019}.
The observed halo gas was likely imprinted by Pop~III supernovae during the galaxy's initial phase of star formation, and subsequently transported outward to its present location via the outflows discussed in Section~\ref{sec5.1}.
The inferred abundance ratios thus reflect the nucleosynthesis of the earliest stellar generations in this system, rather than the instantaneous stellar population of \targ.

\subsection{NIR-dark galaxies as hosts of metal absorbers}
\label{sec5.4}

The heavily dust-obscured nature of \targ\ carries broader implications for survey completeness.
Although recent spectroscopic surveys have revealed a rapidly growing population of intervening metal absorbers at $z>6$ \citep[e.g., ][]{Davies+2023_met_abs, Christensen+2023_qso_abs, Becker+2015_abs}, the galaxy counterparts responsible for enriching these systems remain largely elusive.
At intermediate redshifts ($z\sim2$--$5$), this systematic non-detection has traditionally been attributed to the intrinsic faintness of the hosts \citep{Rhodin+2021_host}.
The unprecedented depth of JWST, however, disfavors this luminosity-limited interpretation at $z\gtrsim7$.
   To quantify this, we place a typical UV-selected galaxy at $z=7.03$. A galaxy with $M_{\rm UV}=-20.5$ has an apparent magnitude of $m_{\rm UV}\approx26.4$ at $\lambda_{\rm obs}\approx1$--1.5\,$\mu$m, and would be detected at ${\rm S/N}\approx30$ in the F115W data at the position of \targ\ ($5\sigma$ depth of 28.4\,mag at the position of \targ; Table~\ref{tab:data}), and at ${\rm S/N}\approx80$ in the deepest NIRCam surveys \citep[$5\sigma$ depths of $\approx29.5$\,mag; e.g.,][]{Eisenstein+2023_JADES}. Even a galaxy two magnitudes fainter ($M_{\rm UV}=-18.5$) would be detected at ${\rm S/N}\approx5$--13. Moreover, if the ${\rm SFR}_{100}\approx26\,M_\odot\,{\rm yr^{-1}}$ of \targ\ were unobscured, it would correspond to $M_{\rm UV}\approx-22.3$ (adopting the UV--SFR conversion of \citealt{Madau+2014} for a Chabrier IMF), i.e., $m_{\rm UV}\approx24.6$, detectable at ${\rm S/N}>100$. The non-detection of \targ\ in F115W ($m>29.0$ at $3\sigma$) therefore implies a rest-frame UV attenuation of $A_{\rm UV}\gtrsim4.3$\,mag, and any host comparable in SFR to typical UV-selected galaxies would be readily detected.
\targ\ suggests an alternative explanation: the hosts of these early metal absorbers may preferentially be heavily dust-obscured.

Dust-obscured galaxies at $z\lesssim5$ have been identified over the past several years \citep{Wang+2019_Hdropout, Barrufet+2023_dark, PerezGonzalez+2023_dark, Alvarez-Marquez+2023_dusty, Xiao+2023_dusty, Bing+2026_dusty}, but such a population is still largely missed at higher redshifts.
In particular, the F356W-to-1.2\,mm flux ratio of \targ\ ($S_{3.6\,\mu\rm m}/S_{1.2\,\rm mm} < 10^{-4}$) is consistent with the ``NIRCam-dark'' starburst galaxies identified and defined by \cite{Sun+25_IRD}.
Nevertheless, the census of these systems has expanded only modestly since the advent of JWST, principally because their intrinsic rest-UV faintness places them at or beyond the sensitivity limits of even the deepest current NIRCam surveys, and their continuum-dominated spectra offer few strong nebular lines for confirmation.
Systematically uncovering and characterizing this hidden population of dust-obscured host galaxies will therefore require large-volume millimeter and submillimeter surveys, which trace the dust continuum and the rest-frame far-infrared cooling lines (e.g., \ciiemi, \oiiiemi) that are largely immune to attenuation, providing a probe complementary to the rest-UV/optical view afforded by JWST.

\section{Summary}
\label{sec6}
We present a multi-wavelength study of a metal-rich circumgalactic medium (CGM) absorber at $z=7.03$ detected in the spectrum of the super-Eddington quasar \tarq\ ($z_{\rm qso}=7.19$), and of its host galaxy \targ, based on JWST/NIRSpec, NIRCam, MIRI, and NOEMA observations. Our main results are:

\begin{enumerate}
    \item In three independent NIRSpec spectra of \tarq, we detect 16 metal absorption transitions from nine ions, from neutral \ion{O}{1} to highly ionized \ion{C}{4}, in a single absorber at $z=7.03$. The absorber has the largest rest-frame equivalent widths observed at $z>6$ ($W_0\approx3.3$\,\AA\ for \ion{Mg}{2}\,$\lambda2796$ and $\approx2.6$\,\AA\ for \ion{C}{2}\,$\lambda1334$, a factor of $\approx2$--4.5 above the strongest previously known $z>6$ absorbers), and its column densities, which are conservative lower limits, lie $\approx1$--1.5\,dex above the median of $z>5$ metal absorbers. It is therefore the strongest metal absorber known at $z>6$.

    \item We identify the host galaxy of the absorber, \targ, which was previously known only as a 1\,mm continuum source, and confirm its redshift spectroscopically for the first time through \oiiiemi, \ha, and \ciiemi\ emission ($z_{\rm sys} = 7.0321 \pm 0.0007$). \targ\ is a heavily dust-obscured, star-forming galaxy with $\log(M_*/M_\odot) = 9.66 \pm 0.24$ and ${\rm SFR}_{100} = 26\pm 10\,M_\odot\,\rm yr^{-1}$ (instantaneous ${\rm SFR}=168\pm35\,M_\odot\,\rm yr^{-1}$). It lies $\approx 2\sigma$ above the star-forming main sequence and follows the high-redshift $L_{\rm [CII]}$--SFR relation, while its SED-based SFRs exceed the dust-corrected \ha\ SFR by factors of $\approx2$--13, indicating that much of its star formation is optically thick even in the rest-frame optical.

    \item At an impact parameter of $\rho = 16\,\rm kpc$, \targ\ is the closest host galaxy known for any absorber at $z>5$, $\approx35\%$ closer than the typical $\rho\approx25$\,kpc of previous pairs. The sightline lies within the virial radius of \targ\ ($r_{\rm vir} \approx 22\,\rm kpc$, or 15--32\,kpc for a $\pm0.5$\,dex uncertainty in halo mass), so it directly probes the inner CGM of a galaxy in the epoch of reionization, a regime rarely probed by previous $z>5$ absorber studies.

    \item With a known host and impact parameter, the absorber can be compared directly with the $W_0$--$\rho$ relations established at $z\sim1$--2. The \ion{Mg}{2}\ and \ion{C}{4}\ equivalent widths agree with these relations at the same physical impact parameter, but exceed them by factors of $\approx5$--20 (\ion{Mg}{2}) and $\approx2$--4.5 (\ion{C}{4}) at the same virial-radius-normalized impact parameter, over the full range of $r_{\rm vir}$ allowed by the halo-mass uncertainty. This suggests that, at a fixed fraction of the virial radius, the CGM at $z\approx7$ contains more metal-enriched gas than at $z\sim1$--2, although a larger sample is needed to confirm this.

    \item Comparing the absorption lines with the systemic redshift of \targ, we find that the absorbing gas is blueshifted by $\lesssim$100\,km\,s$^{-1}$, as expected for an outflow. The outflow velocities range from $\approx50$ to $\approx700$\,km\,s$^{-1}$ depending on the ion (characteristic value $\approx 300\,\rm km\,s^{-1}$), comparable to or above the escape velocity at 16\,kpc ($\approx$220\,km\,s$^{-1}$), so part of the gas may escape the halo. The spread among ions is consistent with a multi-cloud, multiphase outflow, and the travel time of $\gtrsim50$\,Myr places the launch of the gas at $z\gtrsim7.5$, during the active star formation of \targ\ inferred from its SED.

    \item Treating the absorber as a partially covering shell, we estimate a mass outflow rate of $\dot{M}_{\rm out}\approx 64\,M_\odot\,\mathrm{yr^{-1}}$ (plausible range $23$--$260\,M_\odot\,\mathrm{yr^{-1}}$) and a mass loading factor of $\eta\approx 3_{-2}^{+7}$. This constrains the mass loading of a dust-obscured galaxy in the epoch of reionization from absorption by its own CGM. The value is above those of $z\sim1$--2 star-forming galaxies ($\eta\sim0.5$--3) and consistent with recent JWST outflow studies at $z\gtrsim7$, indicating efficient metal transport from the host into its halo.

    \item The abundance ratio $[{\rm Mg/Fe}] = +0.13 \pm 0.05$ is consistent with the mean of absorbers at $z\sim 2$--6 ($+0.37\pm0.35$), although dust depletion of Fe in the CGM of this dust-obscured galaxy may bias it upward. The \ion{C}{2}/\ion{C}{4} ratio ($\log(N_{\rm C\,II}/N_{\rm C\,IV})\leq1.06$) lies within the 0.3\,dex scatter of the empirical redshift trend. Despite column densities $\gtrsim1$\,dex above typical values, the absorber therefore shows a standard ionization balance, and its strong \ion{C}{4} absorption reflects a large total metal column rather than an unusually hard radiation field.

    \item The absorber has $[\mathrm{C/O}] = +0.76^{+0.38}_{-0.31}$ and $[\mathrm{Si/O}] = +0.81^{+0.21}_{-0.18}$. Its [C/O] is the highest among quasar absorbers at $z>5$, $\approx0.85$\,dex above the median of $z\sim5$--6.5 quasar absorbers, and carbon is enhanced by $\approx0.5$--0.7\,dex relative to O, Mg, and Fe. This pattern resembles that of the ISM of UV-bright galaxies at $z\approx7$--9 and is consistent at the $\approx2\sigma$ level with the yields of Population~III core-collapse supernovae and hypernovae, although a purely Population~II origin cannot be excluded. We interpret it as a chemical imprint of the earliest stellar generations carried into the halo by outflows, which extends the known absorbers with Population~III-like abundance patterns from $z\approx3$--6 to $z\approx7$.

    \item \targ\ is undetected in all NIRCam short-wavelength bands (rest-frame $\lambda\lesssim2600$\,\AA), implying $A_{\rm UV}\gtrsim4.3$\,mag, whereas a typical UV-selected galaxy at the same redshift would be detected at ${\rm S/N}\approx30$ in the same data. The hosts of early metal absorbers can therefore be dust-obscured, star-forming galaxies that rest-UV selection misses, and not only intrinsically faint galaxies. This motivates millimeter and submillimeter searches for the hosts of high-redshift absorbers.

\end{enumerate}

The regime revealed by this system, a strong absorber intercepted at only 16 kpc from a heavily dust-obscured, star-forming host, points to two possible next steps.
First, higher-spectral-resolution (e.g., $R\sim 10000$) follow-up will be essential to constrain individual cloud $b$-parameters.
Only such data can break the saturation degeneracy in the column-density estimates and more robustly characterize the underlying nucleosynthetic pattern.
Second, if the enhanced [C/O] observed here is indeed a Pop~III imprint, quasar-absorption spectroscopy toward confirmed high-$z$ galaxy–absorber pairs may prove a complementary route to probing Pop~III enrichment, sidestepping the challenge of directly detecting individual Pop III stellar systems.
Realizing both goals requires a systematic effort to identify high-$z$ absorber hosts even when they are NIR-dark, motivating targeted millimeter and sub-millimeter follow-up of high-$z$ metal-absorber fields with ALMA and NOEMA.

\section*{Acknowledgements}

The authors acknowledge F.~Sun for sharing their data (GO\#7935) in advance and for fruitful discussions.
Q.~F. is deeply grateful to M.~Ouchi for his insightful comments and valuable discussions.
Q.~F. sincerely thanks L.~Pentericci, P.~Dayal, and the local organizing committee of the ``i2i3: still linking galaxy physics from ISM to IGM scales'' conference in Sexten, Italy, as well as all its participants, for their constructive comments and suggestions.
Q.~F. also warmly thanks J.~X.~Prochaska and the scientific and local organizing committees of the ``Celebrating 40 Years of Damped Ly$\alpha$ Systems'' conference in Kashiwa, Japan, together with all its participants, for their helpful feedback and stimulating discussions.
J.A.-M. acknowledges support by grants PID2024-158856NA-I00 \& PIB2021-127718NB-I00 from the Spanish Ministry of Science and Innovation/State Agency of Research MCIN/AEI/10.13039/501100011033 and by “ERDF A way of making Europe”.
M.V. acknowledges financial support by the Independent Research Fund Denmark via grant DFF-8021-00130 and the Carlsberg Foundation via grant CF21-0649.

This work is based in part on observations made with the NASA/ESA/CSA JWST. The data were obtained from the Mikulski Archive for Space Telescopes at the Space Telescope Science Institute, which is operated by the Association of Universities for Research in Astronomy, Inc., under NASA contract NAS 5-03127 for JWST. These observations are associated with program GO 9214 (PIs: C.~Mason \& D.~Stark).
This work is also based on JWST observations associated with programs GO 4762 (PIs: S.~Fujimoto \& G.~Brammer), GO 7935 (PIs: F.~Sun \& X.~Lin), GO 1895 (FRESCO; PI: P.~Oesch), GO 3577 (CONGRESS; PI: E.~Egami), GO 5407 (MEOW; PI: G.~C.~K.~Leung), GTO 1181 (JADES; PI: D.~Eisenstein), GTO 1264 (PI: L.~Colina), and GO 2926 (PI: L.~Colina). We thank the PIs and teams of all these programs for designing and executing the observations and for making the data publicly available.

This work is based on observations carried out under project number W25E004 with the IRAM NOEMA Interferometer. IRAM is supported by INSU/CNRS (France), MPG (Germany) and IGN (Spain).
  All the JWST data used in this paper can be found in MAST: \dataset[doi: 10.17909/v95b-rn02]{https://doi.org/10.17909/v95b-rn02}.

\textit{Facility:} JWST, NOEMA

\textit{Software:} \texttt{grizli} \citep{Brammer+2021_grizli}, \texttt{msaexp} \citep{Brammer2023}, \texttt{Astropy} \citep{Astropy+2022}

\bibliography{sample631}{}
\bibliographystyle{aasjournal}

\appendix
\counterwithin{figure}{section}

\section{Sensitivity of the COG analysis to the assumed Doppler parameter}
\label{app:cog_b}

\begin{figure*}[ht]
    \centering
    \includegraphics[width=\linewidth]{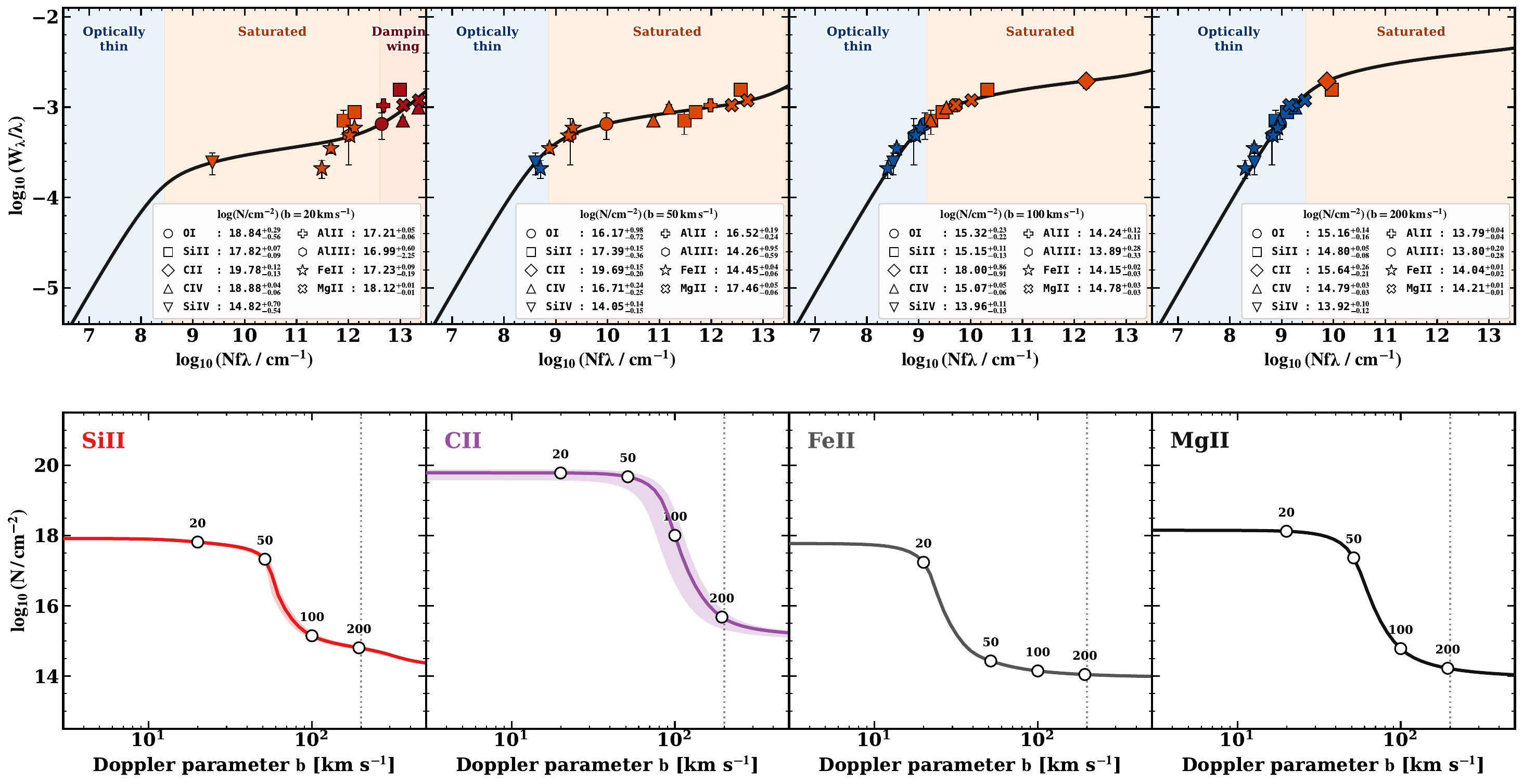}
    \caption{
    Sensitivity of the multi-ion COG analysis to the assumed Doppler parameter $b$.
    Top row (a--d): theoretical COG curves at $b=20, 50, 100,$ and $200\,\rm km\,s^{-1}$, with every measured transition of \targ\ overlaid at its $(\log_{10}(N f\lambda),\log_{10}(W_\lambda/\lambda))$ position using the best-fit column density at that $b$.
    Marker face color encodes the COG regime (linear/linear-to-flat/flat/damped) and marker shape identifies the ion.
    A compact panel-specific legend lists the best-fit $\log N$ (cm$^{-2}$) of every ion at that $b$.
    Bottom row (e--h): $\log_{10}(N/\mathrm{cm}^{-2})$ as a function of $b$ for four representative ions (\siii, \ion{C}{2}, \ion{Fe}{2}, \ion{Mg}{2}).
    Shaded envelopes show the $1\sigma$ Monte Carlo uncertainty.
    White circles mark the four $b$ values shown in the top row, and the vertical dotted line marks the fiducial $b=200\,\rm km\,s^{-1}$ adopted in Table~\ref{tab:absorption}.
    Column densities rise steeply below $b\lesssim 50\,\rm km\,s^{-1}$ as the strong transitions enter the flat and damped branches of the COG.
    }
    \label{fig16:cog_b}
\end{figure*}

The multi-ion COG analysis in Section~\ref{sec3.2} adopts a fixed Doppler parameter $b=200\,\rm km\,s^{-1}$ that is larger than any physically plausible intrinsic $b$, so that the inferred column densities should be conservative lower limits.
In this appendix we quantify how the derived $\log N$ per ion depends on the assumed $b$ value and identify which regime of the COG each measured transition occupies at the fiducial choice.

Figure~\ref{fig16:cog_b} summarizes the analysis in a $4\times 2$ layout.
The top row (panels a--d) presents the theoretical COG curve at $b=20, 50, 100,$ and $200\,\rm km\,s^{-1}$ and, for each $b$, overlays every measured transition at the $(\log_{10}(N f\lambda), \log_{10}(W_\lambda/\lambda))$ position implied by the best-fit column density of its parent ion at \emph{that} $b$.
Each panel carries a compact legend listing the best-fit $\log N$ of every ion at that $b$.
The marker face color encodes which branch of the COG each transition occupies: the optically thin {\it linear} regime ($W\propto N$, blue), the {\it flat} (saturated) plateau ($W\propto \sqrt{\ln \tau_0}$, orange), and the {\it damped} branch ($W\propto \sqrt{N f\gamma}$, red).
At $b=20\,\rm km\,s^{-1}$ most of the strong transitions are firmly on the saturated or damping-wing branches, and the recovered $\log N$ values are correspondingly large.
At $b=200\,\rm km\,s^{-1}$, every transition lies on either the optically thin or the transition to the saturated portion of the COG, exactly as intended by our conservative choice.
Vertical shaded bands in each top panel mark the three canonical regimes of the COG at that $b$ (optically thin -- saturated -- damping wing).

The bottom row (panels e--h) shows $\log_{10}(N/\mathrm{cm}^{-2})$ as a function of $b$ for four representative ions, \siii, \ion{C}{2}, \ion{Fe}{2}, and \ion{Mg}{2}, obtained by refitting the COG at 55 log-spaced values of $b$ between 3 and 500\,\rm km\,s$^{-1}$.
Shaded regions show the $1\sigma$ Monte Carlo uncertainties.
For $b\gtrsim200\,\rm km\,s^{-1}$, the derived column densities are essentially insensitive to $b$, confirming that the linear/linear-to-flat portion of the COG places a firm lower bound on $N$.
Below $b\sim 50\,\rm km\,s^{-1}$, the inferred $\log N$ rises steeply by 1--3 dex as the strongest transitions enter the flat and damped regimes.
White circles on each curve mark the four fiducial values ($b=20, 50, 100, 200\,\rm km\,s^{-1}$) shown in the top row, and the vertical dotted line marks the $b=200\,\rm km\,s^{-1}$ value adopted in Table~\ref{tab:absorption}.

Even at this conservative fiducial, the inferred column densities already exceed the bulk of the high-$z$ metal-absorber literature \citep{Davies+2023_met_abs} by $\approx1$--1.5\,dex (Section~\ref{sec4.1}), confirming that the column densities reported in the main text are conservative lower limits.

\section{Outflow rate estimate}
\label{app:outflow_geom}

\begin{figure}[ht]
    \centering
    \includegraphics[width=0.55\linewidth]{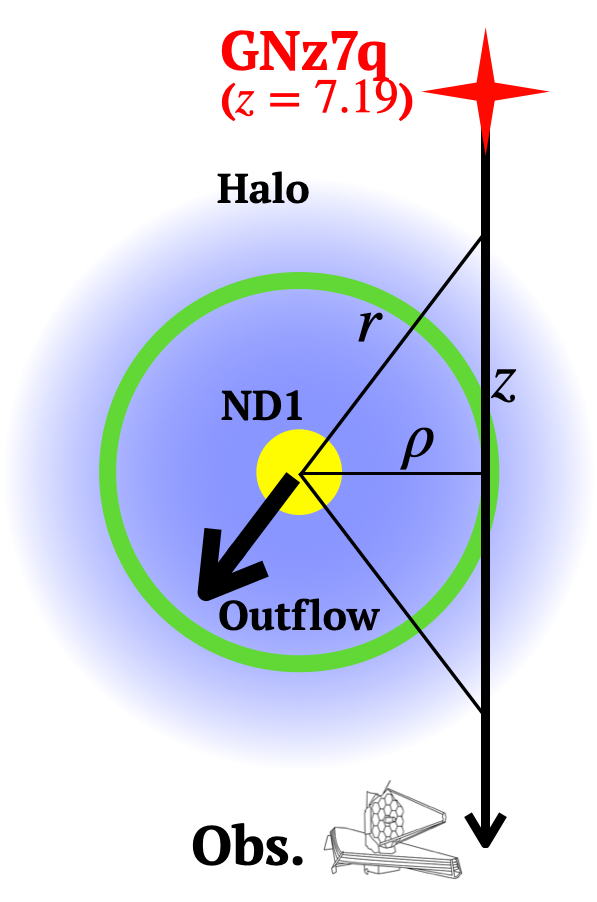}
    \caption{Sketch of the geometry used to convert the absorption measurements into a mass-outflow rate for \targ\ (Section~\ref{sec5.2}).
    The line of sight to the background quasar \tarq\ ($z=7.19$) passes at a projected impact parameter $\rho$ (on-sky separation converted to physical kpc at the host redshift) from the host galaxy \targ\ (yellow).
    The absorbing gas is treated as a partially covering, uniformly enriched thin shell (green circle) of physical radius $r$, embedded in the host halo (light-blue diffuse region). Because we only measure the on-sky separation and the line-of-sight velocity component, both $\rho$ and $v_{\rm out}$ are projected quantities, with $r \geq \rho$ and $|\vec{v}_{\rm out}| \geq v_{\rm out,\,LOS}$; the values quoted in the main text should therefore be interpreted as first-order estimates.}
    \label{fig:app_outflow_geom}
\end{figure}

Figure~\ref{fig:app_outflow_geom} illustrates the geometry underlying the mass-outflow-rate estimate presented in Section~\ref{sec5.2}.
Following \citet{Rupke+2005} and \citet{Chisholm+2017}, we model the metal-enriched gas seen in absorption toward \tarq\ as a spherically symmetric, steady-state outflow of hydrogen number density $n_{\rm H}(r)$ that fills a fraction $C_f$ of the solid angle around the host \targ\ and expands radially outward at velocity $v_{\rm out}$.
Under these assumptions, the instantaneous mass flux crossing any spherical shell of radius $r$ (i.e., the mass outflow rate) centered on \targ\ is:
\begin{align}
    \dot{M}_{\rm out}\equiv \frac{{\rm d} M}{{\rm d} t}=\frac{C_f \cdot \mu m_{\rm H} n_{\rm H} \cdot 4\pi r^2dr}{dt}=C_f \cdot \mu m_{\rm H} n_{\rm H} \cdot 4\pi r^2 v_{\rm out},
    \label{equ: b1}
\end{align}
where $m_{\rm H}$ is the mass of the hydrogen atom and $\mu=1.4$ accounts for the helium contribution to the mean molecular weight, $n_{\rm H}$ is the hydrogen number density at radius $r$, and $v_{\rm out}$ is the radial outflow velocity of the gas at that radius.

The observed hydrogen column density along the line of sight from \targ\ to the background quasar \tarq\ is, by definition, the integral of the local number density along the sight-line coordinate $z$:
\begin{align}
    N_{\rm H} \equiv \int_{-\infty}^{+\infty} n_{\rm H} dz.
    \label{equ: b2}
\end{align}

Solving Equation~\ref{equ: b1} for $n_{\rm H}(r)$ and substituting the result into Equation~\ref{equ: b2}, while using the geometric relation $r^2=\rho^2+z^2$ (see Figure~\ref{fig:app_outflow_geom}) to express $r$ in terms of the on-sky impact parameter $\rho$ and the line-of-sight coordinate $z$, gives:
\begin{align}
    N_{\rm H} = \int_{-\infty}^{+\infty} \frac{\dot{M}_{\rm out}}{C_f \cdot \mu m_{\rm H} \cdot 4\pi r^2 v_{\rm out}}{\rm d}z
    = \frac{\dot{M}_{\rm out}}{C_f \cdot \mu m_{\rm H} \cdot 4\pi v_{\rm out}} \frac{1}{\rho} \int_{-\infty}^{+\infty}\frac{{\rm d}(z/\rho)}{(z/\rho)^2+1}
    =\left.\frac{\dot{M}_{\rm out}}{C_f \cdot \mu m_{\rm H} \cdot 4\pi v_{\rm out}} \frac{1}{\rho} \arctan (z/\rho) \right|_{-\infty}^{+\infty}.
\end{align}

Evaluating the arctan at $z \to \pm\infty$ collapses the integral to $\pi/\rho$, which, when rearranged for $\dot{M}_{\rm out}$, yields the closed-form expression adopted in Section~\ref{sec5.2}:
\begin{align}
    \dot{M}_{\rm out}=4 \mu m_{\rm H} C_f N_{\rm H} \rho v_{\rm out}.
\end{align}
Because $\rho$ is the on-sky projection of the true 3D separation and $v_{\rm out}$ is the line-of-sight component of the outflow velocity, both enter this expression as lower bounds on their intrinsic 3D values. Under the isotropic thin-shell hypothesis, the two projection biases partially cancel in the product $\rho\,v_{\rm out}$, so the derived $\dot{M}_{\rm out}$ remains a physically meaningful first-order estimate. The plausible range $\dot{M}_{\rm out}\sim23$--$260\,M_\odot\,{\rm yr}^{-1}$ quoted in Section~\ref{sec5.2} reflects the propagation of the residual systematic uncertainties on $C_f$, $[{\rm Si/H}]$, and $v_{\rm out}$ through this expression, holding the geometry of Figure~\ref{fig:app_outflow_geom} fixed.

\end{document}